\documentclass[nofootinbib,superscriptaddress,showpacs,eqsecnum,tightenlines,11pt]{revtex4}

\usepackage{amsfonts}
\usepackage{amsmath}
\usepackage{amssymb}  
\usepackage{hyperref}

\newcommand{\gam}{{}^{\text{\tiny{($\gamma$)}}}\!}
\newcommand{\bst}{{}^{\tiny{\mathcal{B}}}\!\!}
\newcommand{\rot}{{}^{\tiny{\mathcal{R}}}\!\!}
\newcommand{\rotE}{{}^{\tiny{\mathcal{R}}}\mathcal{E}}
\newcommand{\bstE}{{}^{\tiny{\mathcal{B}}}\mathcal{E}}
\newcommand{\brA}{{}^{\tiny{\mathcal{B},\mathcal{R}}}\!\!\mathcal{A}}
\newcommand{\rbA}{{}^{\tiny{\mathcal{R},\mathcal{B}}}\!\!\mathcal{A}}

\newcommand{\ub}{\underline{u}\,}
\newcommand{\vb}{\underline{v}\,}
\newcommand{\Yb}{\underline{Y}}
\newcommand{\SU}{\text{SU}}
\newcommand{\SO}{\text{SO}}
\newcommand{\SL}{\text{SL}}
\newcommand{\su}{\mathfrak{su}}
\newcommand{\so}{\mathfrak{so}}
\newcommand{\sll}{\mathfrak{sl}}
\newcommand{\ch}{\text{ch}\,\theta}
\newcommand{\sh}{\text{sh}\,\theta}
\newcommand{\htan}{\text{th}\,\alpha}
\newcommand{\tra}{^{\text{\tiny{T}}}}

\def\be{\begin{equation}}
\def\ee{\end{equation}}
\def\ba{\begin{eqnarray}}
\def\ea{\end{eqnarray}}
\def\f{\frac}

\begin{document}
\title{A new look at Lorentz-Covariant Loop Quantum Gravity}
\author{Marc Geiller}
\email{mgeiller@apc.univ-paris7.fr}
\author{Marc Lachi\`eze-Rey}
\email{mlr@apc.univ-paris7.fr}
\affiliation{Laboratoire APC - Astroparticule et Cosmologie, Universit\'e Paris Diderot Paris 7, 75013 Paris, France}
\author{Karim Noui}
\email{karim.noui@lmpt.univ-tours.fr}
\affiliation{Laboratoire de Math\'ematiques et Physique Th\'eorique, Universit\'e Fran\c cois Rabelais, Parc de Grandmont, 37200 Tours, France}


\begin{abstract}
In this work, we study the classical and quantum properties of the unique commutative Lorentz-covariant connection for loop quantum gravity. This connection has been found after solving the second-class constraints inherited from the canonical analysis of the Holst action without the time-gauge. We show that it has the property of lying in the conjugacy class of a pure $\su(2)$ connection, a result which enables one to construct the kinematical Hilbert space of the Lorentz-covariant theory in terms of the usual $\SU(2)$ spin-network states. Furthermore, we show that there is a unique Lorentz-covariant electric field, up to trivial and natural equivalence relations. The Lorentz-covariant electric field transforms under the adjoint action of the Lorentz group, and the associated Casimir operators are shown to be proportional to the area density. This gives a very interesting algebraic interpretation of the area. Finally, we show that the action of the surface operator on the Lorentz-covariant holonomies reproduces exactly the usual discrete $\SU(2)$ spectrum of time-gauge loop quantum gravity. In other words, the use of the time-gauge does not introduce anomalies in the quantum theory.
\end{abstract}

\maketitle

\section{Introduction}

\noindent Canonical loop quantum gravity \cite{al,rovelli,thiemann} is an attempt to quantize general relativity as a constrained Hamiltonian system, which takes as a starting point the first-order formulation of gravity in terms of connection and triad variables. The historical motivation for this choice was the observation by Sen \cite{sen} that working with a self-dual connection (which is a complex $\sll(2,\mathbb{C})$ connection in the Lorentzian case) leads to a canonical theory which is free of second-class constraints, and features a Hamiltonian which is polynomial in the basic variables. Ashtekar \cite{ashtekar-variables} showed later on that the complex formulation of Sen can be reached from a canonical transformation on the phase space of general relativity, and investigated the canonical structure as well as the Poisson algebra of the theory. This work has opened an important road toward the canonical quantization of general relativity, as it constitutes the starting point of loop quantum gravity. However, it was soon realized that the reality conditions which are needed in order to recover the real phase space of general relativity are complicated to deal with at the quantum level. Up to now, no one knows how to impose properly these conditions in the quantum theory. Barbero \cite{barbero} and Immirzi \cite{immirzi} have found a very nice alternative to address this problem. They suggested to work with a real $\su(2)$ connection, the famous Ashtekar-Barbero connection, which can be obtained using a one-parameter family of canonical transformations on the complex phase space of the Ashtekar formulation of general relativity. In other words, such a canonical transformation makes the phase space real, and the fundamental configuration variable becomes a real $\su(2)$ connection which depends on the so-called Barbero-Immirzi parameter labeling the canonical transformations. At the classical level, this parameter is irrelevant if the matter Lagrangian does not contain fermions, and simply drops out of the equations of motion by virtue of Bianchi identities. However, at the quantum level, it appears explicitly  in the spectra of geometric operators \cite{ashtekar-lewandowski,rovelli-smolin}, and in the black hole entropy formula \cite{rovelli-black hole,ABK,meissner,agullo,ENP}. The status of this free parameter is still quite unclear, and many people have questioned its physical relevance \cite{rovelli-thiemann}.

Even if the introduction of the Ashtekar-Barbero connection has eliminated the question of the reality conditions, it has raised new problems. Let us mention two of them. The first one concerns the exact physical significance of the Barbero-Immirzi parameter, and the second one the Lorentz covariance of the Ashtekar-Barbero connection \cite{samuel}. Indeed, the canonical transformation introduced by Barbero transforms the initial complex  $\sll(2,\mathbb{C})$ connection into a real $\su(2)$ connection and, at the same time, turns the complex $\sll(2,\mathbb{C})$ symmetry algebra into an $\su(2)$ symmetry algebra. As a consequence, working with the Ashtekar-Barbero connection can be done only at the expense of losing the Lorentz covariance of the theory.

It is therefore natural to ask whether it is possible to find a Lorentz-covariant generalization of the Ashtekar-Barbero connection. This question is also very important because many central results of loop quantum gravity, like the discrete spectra of the geometrical operators, depend crucially on the fact that the symmetry algebra of the theory is $\su(2)$ and not $\so(3,1)$. What happens if one tries to quantize the manifestly Lorentz-invariant theory, i.e. without making any gauge choice at all? In particular, one can ask whether the predicted discreteness of the geometric operators \cite{ashtekar-lewandowski,rovelli-smolin} is an artifact of this gauge fixing or not. Indeed, it is \textit{a priori} not completely obvious that the discreteness of the quantum geometry at the kinematical level survives the presence a full Lorentz symmetry algebra. These aspects have already been discussed in a remarkable series of articles by Alexandrov and collaborators \cite{alexandrov1,alexandrov2,alexandrov3,alexandrov4}, and the purpose of this paper is to give a new look at them.

Our starting point is the Holst action \cite{holst}, which is a first-order action for gravity leading to a canonical theory expressed in terms of the $\su(2)$ Ashtekar-Barbero connection. More precisely, the Holst action is by construction invariant under the Lorentz algebra, and to derive the canonical structure of $\SU(2)$ loop gravity starting from this action, one usually works in the so-called time-gauge. This choice corresponds to fixing the boost part of the action, thereby reducing the internal $\so(3,1)$ symmetry algebra to its rotational $\su(2)$ subalgebra. What happens if one does not gauge fix the theory and chooses to work with the full Lorentz symmetry? Naturally, one expects to find from the canonical analysis a candidate for a Lorentz generalization of the Ashtekar-Barbero connection. However, the Hamiltonian analysis of the Lorentz-covariant Holst action is technically quite involved, and much harder to perform when one does not consider the time-gauge. Furthermore, a natural Lorentz generalization of the Ashtekar-Barbero connection does not arise naturally from the canonical analysis, as opposed to what happens for the $\su(2)$ connection
in the time-gauge. The main reason is that second-class constraints are present in the canonical analysis, revealing the fact that some of the degrees of freedom necessary to write an action principle are redundant. Apart from being nonphysical, these extra degrees of freedom are also not pure gauge, and they have to be removed for example by solving explicitly the second-class constraints \cite{henneaux-teitelboim}. In general, there are many different (but equivalent) parametrizations of the space of solutions to the second-class constraints, each leading to a particular parametrization of the physical phase space. These distinct parametrizations can potentially lead to different generalizations of the Ashtekar-Barbero connections, all of them being of course related to one another.

To our knowledge, three different roads have been explored to handle the second-class constraints inherited from the canonical analysis of the Holst action without any gauge fixing \cite{alexandrov4,cianfrani-montani,barros}. They lead to three different descriptions of the physical phase space.

Alexandrov did not solve the second-class constraints explicitly, but computed instead the Dirac bracket \cite{henneaux-teitelboim}. He has found a two-parameter family of Lorentz connections, but the drawback of his approach is that the associated symplectic structure has a very complicated form. Indeed, the fundamental variables are a connection and its conjugate momentum, and the connection components are generically noncommuting variables. With a proper choice of the parameters entering the definition of this family of Lorentz-connections, it is possible to extract the unique commutative connection \cite{alexandrov4}, which however does not transform properly under timelike diffeomorphisms. To obtain the right transformation property \cite{alexandrov5}, another choice is possible for the two parameters, but in this case the connection is again noncommutative. The quantization of the noncommutative theory is very difficult to perform and so far no representation of the associated quantum algebra does exists. If one chooses to work with the commutative connection, the framework of Alexandrov is still quite complicated because it makes use of Dirac brackets which are not easy to handle. This is our motivation for solving the second-class constraints before building the commutative connection. Indeed, we will show that this simplifies the algebraic structure of the theory, and that it enables one to study the properties of the connection in greater details.

Barros e S\'a was the first to solve the constraints explicitly and to give a parametrization of the phase space with a very simple symplectic structure and only first-class constraints. The problem of his approach, as emphasized by Alexandrov, is that this formalism is \textit{a priori} not well suited to study Lorentz-covariant canonical gravity, because although the generators of the full Lorentz symmetry are still present, the Lorentz-covariant connection does not appear explicitly.

Cianfrani and Montani have proposed another resolution of the second-class constraints with, at the end, another parametrization of the physical phase space and a link with the work of Alexandrov.

This paper continues the work initiated in \cite{GLNS}, and provides the detailed proof of the existence and uniqueness of the commutative Lorentz connection, together with the study of the quantum theory. It is based on the analysis of Barros e S\'a, who followed the ideas of Peldan \cite{peldan} to show that it is possible to solve the second-class appearing in the Holst action without the time-gauge. The resulting phase space is parametrized by two pairs of canonical variables, $(A,E)$ and $(\chi,\zeta)$, satisfying a set of 10 first-class constraints. The first pair is constituted by an $\su(2)$ generalization of the Ashtekar-Barbero connection and its conjugate momentum, and the second pair consists of two vectors with values in $\mathbb{R}^3$. As a consequence, there is \textit{a priori} no natural Lorentz connection in this phase space formulation. However, the theory is clearly invariant the Lorentz algebra, because six out of the ten first-class constraints generate the symmetries associated to the action of the Lorentz algebra. The four remaining constraints are the diffeomorphisms and the scalar constraints. Guided by the action of the Lorentz constraints on the fundamental variables, we found in \cite{GLNS} that there is a Lorentz connection hidden in the phase space given by Barros e S\'a. More precisely, we showed that it is possible to construct an unique Lorentz connection which depends on the variables $A$ and $\chi$ only. It is important to mention that our connection should be equivalent to the commutative connection found in \cite{alexandrov4} by Alexandrov and Livine, even if we did not show the link explicitly. Our approach and that of Alexandrov are of course equivalent, the only difference being in the technical aspects. However, for reasons that will become explicit in the core of this paper, it seems to us that the present approach allows one to study the classical and quantum theories in a much simpler and transparent way. In this work, we also show that the unique commutative Lorentz connection is gauge equivalent to a pure $\su(2)$ connection. More precisely, there exists a pure boost $B$ which maps the covariant connection to an $\su(2)$ connection. This boost is in fact the one which maps the parameter $\chi$ to zero. Then, we construct the Lorentz-covariant ``electric field'' which  also appears to be unique up to a  natural equivalence relation. This electric field allows us to construct a Lorentz-covariant area of surfaces. The quantization of the theory in this framework turns out to be straightforward because of the simple form of the symplectic structure. Surprisingly, it is also possible to construct completely the kinematical Hilbert space and to compute the spectrum of the area operator. Physical states are \textit{a priori} described in terms of Lorentz spin-networks, but they reduce to $\su(2)$ spin-networks because of the property that we have mentioned above. Therefore, the physical Hilbert space is totally well-defined and there is no problem of divergences due to the noncompactness of the Lorentz group. We show that the area operator is Lorentz-invariant, and generalizes the standard area operator used in the time-gauge Ashtekar-Barbero theory. The action of this area operator on Lorentz spin-network states gives a spectrum which turns out to be the one obtained in the time-gauge. As a consequence, the discreteness of the area operator at the kinematical level in loop quantum gravity is not an artifact of the time-gauge. In other words, starting with Lorentz-invariant spin-network states and a Lorentz-invariant area operator leads to an $\su(2)$ like spectrum for the area operator at the kinematical level.

The outline of this work is the following. In the second section, we briefly recall the canonical analysis of first-order gravity with nonvanishing Barbero-Immirzi parameter. In particular, we give the expression of the constraints and the phase space variables once the second-class constraints have been solved. In the third section, we propose a physical interpretation for the variable $\chi$, which allows us to introduce new variables with a clear geometrical meaning, and in terms of which we can demonstrate the existence of a unique one-form which is covariant under the action of the Lorentz group. In the fourth section, we show that it is possible to find a Lorentz boost which sends the unique Lorentz-covariant connection to a pure $\su(2)$ connection. Then, we use the same procedure we have used to construct the Lorentz-covariant connection, to find the covariant electric field. The Casimir operators associated to the electric field define kinematical observables and, surprisingly, we show that the two Casimir operators are proportional to the local Lorentz-invariant area. Finally, we construct the quantum theory and the Lorentz-invariant area operator of loop quantum gravity, and show that it acts on kinematical states to give the usual discrete $\SU(2)$ spectrum.

Notations are such that $\mu,\nu,\dots$ refer to spacetime indices, $a,b,\dots$ to spatial indices, $I,J,\dots$ to $\so(3,1)$ indices, and $i,j,\dots$ to $\su(2)$ indices. We will assume that the four-dimensional spacetime manifold $\mathcal{M}$ is topologically $\Sigma\times\mathbb{R}$, where $\Sigma$ is a three-dimensional manifold without boundaries.

\section{First order Lorentz-covariant gravity}

\noindent In this section, we recall the main steps of the Hamiltonian analysis of the Lorentz-invariant Holst action without gauge fixing. In particular, we recall the structure of the algebra of constraints, and give the explicit solution to the second-class constraints, as well as the parametrization of the phase space that will serve as a starting point to build the Lorentz-covariant connection. This section relies mostly on the paper of Barros e S\'a \cite{barros}. 

\subsection{Hamiltonian formulation}

\noindent Our starting point for this analysis is the Holst action \cite{holst}, a generalized Hilbert-Palatini first-order action with nonvanishing Barbero-Immirzi parameter. In terms of the cotetrad $e^I_\alpha(x)$ and the Lorentz connection one-form $\omega^{IJ}_\alpha(x)$, the
associated Lagrangian density is given by
\be\nonumber
\mathcal{L}[e,\omega]=e^I\wedge e^J\wedge\star\gam F_{IJ},
\ee
where $F[\omega]=d\omega+\omega\wedge\omega$ is the curvature two-form of the connection $\omega$, and $\star$ is the usual Hodge dual operator in the Lie algebra $\so(3,1)$, whose definition is recalled in appendix \ref{lorentz algebra}. For each $\gamma\neq0$, we have defined the endomorphism
\ba
\so(3,1)&\longrightarrow&\so(3,1)\nonumber\\
\xi&\longmapsto&\gam\xi=(1-\gamma^{-1}\star)\xi.\nonumber
\ea
This map is invertible provided that $\gamma^2\neq-1$, and its inverse is given by the relation
\be\nonumber
\xi=\f{\gamma^2}{1+\gamma^2}(1+\gamma^{-1}\star)\gam\xi.
\ee
When $\gamma^2=-1$, this map is a projection onto the self-dual or anti self-dual subalgebra of $\so(3,1)$. It is well known that the Holst action is equivalent to the Hilbert-Palatini action, and, if the cotetrad is not degenerate (i.e. if its determinant is different from zero), it is equivalent to the Einstein-Hilbert action and the Barbero-Immirzi parameter $\gamma$ drops out of the action by virtue of the Bianchi identities.

Now, we are interested in the canonical analysis of the Holst Lagrangian. For this purpose, we perform a $3+1$ decomposition on the temporal and spatial indices. A few calculations lead to the following canonical expression:
\ba
\mathcal{L}[e,\omega]
&=&\gam\pi^a_{IJ}\dot{\omega}^{IJ}_a-g^{IJ}\mathcal{G}_{IJ}-N\mathcal{H}-N^a\mathcal{H}_a,\label{lagrangian}
\ea
where we have introduced the Lagrange multiplier $g^{IJ}\equiv-\omega^{IJ}_0$, the lapse $N$, and the shift $N^a$, enforcing respectively the Gauss, Hamiltonian, and diffeomorphism constraints
\be\label{constraints}
\mathcal{G}_{IJ}=D_a\gam\pi^a_{IJ},\qquad\qquad
\mathcal{H}=\pi^a_{IK}\pi^{bK}_J\gam F^{IJ}_{ab},\qquad\qquad
\mathcal{H}_a=\pi^b_{IJ}\gam F^{IJ}_{ab}.
\ee
As usual, $\dot{\omega}=\partial_0\omega$ holds for the time derivative of $\omega$.
These constraints are expressed in terms of the spatial connection components $\omega_a^{IJ}$, and the canonical momenta defined by
\be\nonumber
\pi^a_{IJ}\equiv\f{\delta\mathcal{L}}{\delta\gam\dot{\omega}^{IJ}_a}=\epsilon_{IJKL}\epsilon^{abc}e^K_be^L_c.
\ee
Since $\pi^a_{IJ}=-\pi^a_{JI}$ has 18 components, and the cotetrad has only 12 independent components, we need to impose 6 primary constraints of the form
\be\label{simplicity}
\mathcal{C}^{ab}=\epsilon^{IJKL}\pi^a_{IJ}\pi^b_{KL}\approx0,
\ee
in order to parametrize the space of momenta in terms of the variables $\pi^a_{IJ}$ instead of the cotetrad. As a result, classically, it is equivalent to work with the 12 components $e_a^I$, or with the 18 components $\pi^a_{IJ}$ constrained by the 6 equations $\mathcal{C}^{ab}\approx0$. As a consequence of this procedure, the nonphysical Hamiltonian phase space is now parametrized by the pair of canonically conjugated variables $(\omega_a^{IJ},\pi^a_{IJ})$, with the set of constraints (\ref{constraints}) to which we add  $\mathcal{C}^{ab}\approx0$.

We can now write down the expression of the total Hamiltonian,
\be\label{hamiltonian}
\mathcal{H}_{tot}\equiv\pi^a_{IJ}\gam\dot{\omega}^{IJ}_a-\mathcal{L}=g^{IJ}\mathcal{G}_{IJ}+N\mathcal{H}+N^a\mathcal{H}_a+c_{ab}\mathcal{C}^{ab}
\ee
where $c_{ab}=-c_{ba}$ are Lagrange multipliers. The Poisson bracket is given by
\be\nonumber
\big\{\omega^{IJ}_a(x),\gam\pi^b_{KL}(y)\big\}=\big\{\gam\omega^{IJ}_a(x),\pi^b_{KL}(y)\big\}=\delta^b_a(\delta^I_K\delta^J_L-\delta^J_K\delta^I_L)\delta^3(x-y).
\ee
Since we choose to work with the fundamental variables $(\gam\omega,\pi)$, we need to express the constraints in terms
of these variables uniquely. To do so, it is sufficient to give the expression of the curvature $F[\omega]$ in terms of
$\gam\omega$, that is
\be\nonumber
\gam F^{IJ}_{ab}=\partial_{[a}\gam\omega^{IJ}_{b]}+\f{\gamma ^2}{1+ \gamma ^2}\left[\gam\omega^I_{[aK}\gam\omega^{KJ}_{b]}+\f{1}{\gamma}\star\left(\gam\omega^I_{[aK}\gam\omega^{KJ}_{b]}\right)\right].
\ee
This closes the first step of the canonical analysis, namely, the description of the fundamental variables, and the definition of the Poisson bracket and the primary constraints. The next step is the constraint analysis.

\subsection{Constraint analysis}

\noindent The constraint analysis of the Hamiltonian (\ref{hamiltonian}) is rather standard, and has been performed for the Hilbert-Palatini and the Holst action respectively in \cite{peldan} and \cite{barros}. Here we will not reproduce all the steps, but only focus on the second-class constraints and their resolution.

The Hamiltonian theory that we have constructed in the previous subsection is described by the 18 components $\omega_a^{IJ}$ of the $\so(3,1)$ connection, and the 18 components $\pi^a_{IJ}$. These variables are constrained to satisfy the 16 algebraic relations (\ref{constraints}) and (\ref{simplicity}). It is therefore straightforward to realize that in order to recover the $4$ phase space degrees of freedom (per spacetime points) of gravity, the theory needs to have secondary constraints, which in addition have to be second-class. This is indeed the case. To understand how this comes about, notice that the algebra of constraints fails to close because the scalar constraint $\mathcal{H}$ does not commute weakly with the simplicity constraint $\mathcal{C}^{ab}$. In fact, their Poisson bracket generates the 6 additional secondary constraints 
\be\nonumber
\mathcal{D}^{ab}=\star\pi^c_{IJ}\left(\pi^{aIK}D_c\pi^{bJ}_{~~K}+\pi^{bIK}D_c\pi^{aJ}_{~~K}\right)\approx0.
\ee
After this step, the Dirac algorithm closes, and there are no tertiary constraints \cite{peldan,alexandrov2,barros}.

Now, one has to make the separation between first-class and second-class constraints. To do so, one computes the Dirac matrix, whose components are given by the Poisson brackets $\big\{\phi_1,\phi_2\big\}$ between any pair of constraints $(\phi_1,\phi_2)$. The dimension of this matrix is $22\times22$, but it is rather immediate to show that its kernel is 10-dimensional. Physically, this means that among the 22 constraints $\mathcal{H}$, $\mathcal{H}_a$, $\mathcal{G}_{IJ}$, $\mathcal{C}^{ab}$, and $\mathcal{D}^{ab}$, 10 are first-class, and the remaining 12 are second-class. Moreover, one can check that $\mathcal{C}^{ab}\approx0$ and $\mathcal{D}^{ab}\approx0$ form a set of second-class constraints. Since the constraints (\ref{constraints}) are first-class (up to the previous second-class constraints), they generate the symmetries of the theory, namely, the spacetime diffeomorphisms and the Lorentz gauge symmetry. We have now 18 connection components and 18 conjugate momenta, the 10 first-class constraints $\mathcal{G}^{IJ}$, $\mathcal{H}$, and $\mathcal{H}_a$ (generating 10 additional gauge symmetries), and the 12 second-class constraints $\mathcal{C}^{ab}$ and $\mathcal{D}^{ab}$. We are indeed left with 4 phase space degrees of freedom per spatial point.

\subsection{Solving the second-class constraints}

\noindent Now that we have clarified the canonical structure of the theory and identified the second-class constraints, we have to solve them explicitly, or implement them in the symplectic structure by using the Dirac bracket. Our point of view is that the formalism is more transparent and close to the physical intuition if we solve the second-class constraints instead of computing the Dirac bracket.

Following the idea of Peldan \cite{peldan}, Barros e S\'a solves the constraints by setting
\be\label{solution}
\pi^a_{0i}=\f{1}{2}E^a_i,\qquad\qquad\pi^a_{ij}=\f{1}{2}E^a_{[i}\chi_{j]},
\ee
where $\chi^i=-e^i_0/e^0_0$ encodes the deviation of the normal to the hypersurfaces from the time direction \cite{barros}. In the time-gauge, this field is set to zero, and the field $E^a_i$ corresponds to the usual densitized triad of loop gravity. If we plug the solution (\ref{solution}) into the canonical term of the Lagrangian (\ref{lagrangian}), the upshot is
\be\nonumber
\gam\pi^a_{IJ}\dot{\omega}^{IJ}_a=E^a_i\dot{A}^i_a+\zeta^i\dot{\chi}_i,
\ee
where
\be\label{new variables}
A^i_a=\gam\omega^{0i}_a+\gam\omega^{ij}_a\chi_j,\qquad\qquad\zeta^i=\gam\omega^{ij}_aE^a_j.
\ee
This shows that the phase space can be parametrized by the two pairs of canonical variables $(A^i_a,E^a_i)$ and $(\chi_i,\zeta^i)$, their Poisson bracket being given by
\be\label{phase space}
\big\{A^i_a(x),E^b_j(y)\big\}=\delta^i_j\delta^b_a\delta^3(x-y),\qquad\qquad
\big\{\chi_i(x),\zeta^j(y)\big\}=\delta^j_i\delta^3(x-y).
\ee
As a remark, notice that if we work in the time-gauge (i.e. set $\chi=0$), the variable $$A^i_a=\gam\omega^{0i}_a=(1-\gamma^{-1}\star)\omega^{0i}_a$$ 
coincides with the standard Ashtekar-Barbero connection.

The last step is now to express the first-class constraints (\ref{constraints}) in terms of the new phase space variables (\ref{phase space}). To do so, one has to invert the relation (\ref{new variables}), which can be done by writing
\be\nonumber
\gam\omega^{0i}_a=A^i_a-\gam\omega^{ij}_a\chi_j,\qquad\qquad\gam\omega^{ij}_a=\f{1}{2}\left(Q^{ij}_a-E_a^{[i}\zeta^{j]}\right),
\ee
where $E^i_a$ is the inverse of $E^a_i$, and $Q_a^{ij}=Q_a^{ji}$ has a vanishing action on $E^a_i$. Barros e S\'a has derived the explicit expression of the remaining constraints in terms of the variables (\ref{phase space}). As we only need the diffeomorphism and Lorentz constraints in this work, we will not give the expression of the Hamiltonian constraint, which is rather complicated.

To give the expression of these constraints, we introduce a vectorial notation which consists in identifying the variables
of the type $X^i$ with vectors in $\mathbb R^3$. The contraction of internal Lie algebra indices reduces to scalar products
in $\mathbb R^3$. As a result, the vector constraint $\mathcal{H}_a$ takes the smeared form
\ba
\mathcal{H}_a(N^a)&=&\int d^3x\,N^a\Bigg[E^b\cdot\partial_{[a}A_{b]}+\zeta\cdot\partial_a\chi+\f{\gamma^2}{1+\gamma^2}\Big[(E^b\cdot A_b)(A_a\cdot\chi)-(E^b\cdot A_a)(A_b\cdot\chi)\nonumber\\
&&\qquad\qquad\qquad+(A_a\cdot\chi)(\zeta\cdot\chi)-(A_a\cdot\zeta)+\f{1}{\gamma}\left(E^b\cdot(A_b\wedge A_a)+\zeta\cdot(\chi\wedge A_a)\right)\Big]\Bigg].\nonumber\\\label{vector constraint}
\ea
The Lorentz constraint $\mathcal{G}_{IJ}$ can be split between its boost part $\mathcal{B}_i\equiv\mathcal{G}_{0i}$, and its rotational part $\displaystyle\mathcal{R}_i\equiv(1/2)\epsilon_i^{~jk}\mathcal{G}_{jk}$. These two constraints are given by
\begin{subequations}\label{gauss}
\ba
\mathcal{B}(u)&=&\int d^3x\left[-\partial_au\cdot\left(E^a-\f{1}{\gamma}\chi\wedge E^a\right)-u\cdot(\chi\wedge E^a)\wedge A_a-u\cdot\zeta+(\zeta\cdot\chi)(\chi\cdot u)\right],\\
\mathcal{R}(v)&=&\int d^3x\left[\partial_av\cdot\left(\chi\wedge E^a+\f{1}{\gamma}E^a\right)+v\cdot(A_a\wedge E^a-\zeta\wedge\chi)\right].
\ea
\end{subequations}
At this level, one can check for consistency that this is indeed the Lorentz algebra, since we have the relations (see appendix \ref{appendixA})
\ba
\big\{\mathcal{B}(u_1),\mathcal{B}(u_2)\big\}&=&-\mathcal{R}(u_1\wedge u_2),\nonumber\\
\big\{\mathcal{R}(v_1),\mathcal{R}(v_2)\big\}&=&\mathcal{R}(v_1\wedge v_2),\nonumber\\
\big\{\mathcal{B}(u_1),\mathcal{R}(v_2)\big\}&=&\mathcal{B}(u_1\wedge v_2).\nonumber
\ea
Now our task is to use the phase space variables (\ref{phase space}) to build a connection that transforms properly under the boosts and rotations (\ref{gauss}).

Before going into further details, let us finish this section by quickly analyzing what happens when the time-gauge is imposed. To work with the time-gauge, we impose the condition $\chi\approx0$, which has to be interpreted as a new constraint in the theory. It is clear that this constraint, together with the boost generator, form a set of second-class constraints, that we can solve explicitly by taking $\chi=0$ and $\zeta=\partial_aE^a$. By doing so, the variables $(\chi,\zeta)$ are eliminated from the theory, and we see from the expression (\ref{new variables}) that $A$ becomes the standard Ashtekar-Barbero connection. In other words, we recover the usual $\su(2)$ theory.

\section{Lorentz-covariant connection}
\label{covariant connection}

\noindent Our goal is to build a Lorentz-covariant connection $\mathcal{A}$ which depends only on the variables $A$ and $\chi$. There are two natural reasons to make such a choice. The first one is technical, as we would like to build a connection which is commutative with respect to the Poisson bracket, in order to have a chance to construct at least the kinematical Hilbert space following the techniques of usual loop quantum gravity. The second reason is more conceptual, and relies on the fact that it is possible to interpret $\chi$ as a connection variable instead of a cotetrad component. The argument to see how $\chi$ can be interpreted as a Lorentz connection is very simple, and is based on the Lorentz invariance of the first-order Hilbert-Palatini action $S[e,\omega]$. Indeed, this invariance allows one to establish the trivial identity
\be\label{discussion on B}
S[e,\omega]=S[B(\chi)\tilde{e},\omega]=S[\tilde{e},\tilde{\omega}],
\ee
where the action $S[e,\omega]$ is expressed in terms of the non time-gauge cotetrad $e$, and the Lorentz connection $\omega$, and the action $S[\tilde{e},\tilde{\omega}]$ is expressed in terms of the time-gauge cotetrad $\tilde{e}$, and the boosted connection 
\be\nonumber
\tilde{\omega}\equiv B\triangleright\omega=B^{-1}\omega B+B^{-1}dB.
\ee
Here, $B \equiv B(\chi)$ is the unique boost which sends $e$ to $\tilde{e}$. It is clear that $B$ depends only on the variable $\chi$. Because of identity (\ref{discussion on B}), it is equivalent to interpret $\chi$ as a cotetrad component, or as a connection variable entering the definition of the Hilbert-Palatini action $S[\tilde{e},\tilde{\omega}]$. This argument has to be clarified, and this is exactly what we are going to do in this section. 

More precisely, we are going to recall the status of the phase space variables that we are left with after solving the second-class constraints with the procedure of Barros e S\'a. In the first subsection, we will clearly show how $\chi$ can be interpreted as a boost parameter, as we expect from the previous argument. In the second subsection, we are going to suggest a general form for the Lorentz connection, and then show how it can be uniquely determined with the requirement of covariance under the action of the rotations and the boosts. In fact, we will establish an even stronger statement, by showing that the condition of covariance under the boosts is enough to determine uniquely the Lorentz-covariant connection, which is then automatically transformed in a covariant manner under the action of the rotations.

\subsection{Basic variables}

\noindent Two natural pairs of canonical variables, $(A,E)$ and $(\chi,\zeta)$, have emerged from the canonical analysis performed by Barros e S\'a. In this section, we will be interested only in the connection variables $A$ and $\chi$, which are chosen to be the configuration variables, and we will postpone the discussion concerning the momentum variables $E$ and $\zeta$ to section \ref{moments}.

In order to understand the physical interpretation of the configuration variables $A$ and $\chi$, let us first look at their transformation under the action of the boosts and rotations. Using the results given in appendix \ref{formulas}, we have
\ba
\big\{\mathcal{B}(u),A\big\}&=&du-\f{1}{\gamma}du\wedge\chi-(u\wedge A)\wedge\chi,\nonumber\\
\big\{\mathcal{B}(u),\chi\big\}&=&u-(u\cdot\chi)\chi,\nonumber\\
\big\{\mathcal{R}(v),A\big\}&=&\chi\wedge dv-\f{1}{\gamma}dv+A\wedge v,\nonumber\\
\big\{\mathcal{R}(v),\chi\big\}&=&\chi\wedge v.\nonumber
\ea
If the time-gauge is not imposed, one cannot interpret the connection variable $A$ as the rotational component of a Lorentz connection. In fact, it is not even clear that one could find an interesting physical interpretation for this variable. On the contrary, it seems that although $A$ is a good technical variable to solve the second-class constraints, and to describe the physical phase space of gravity without second-class constraints, it is not a good physical variable out of which one could extract physical informations. In the next subsections, we are going to show what the ``good" and unique physical connection is.

Likewise, the status of the variable $\chi$ is \textit{a priori} quite unclear. The main reason for this is that it transforms in a nonlinear way under the action of boosts. This makes the geometrical interpretation of $\chi$ rather obscure, and leads to technical difficulties when handling the algebra of constraints. In fact, this difficulty can be solved quite easily because, up to a normalization factor that we are going to precise right after, $\chi$ can be interpreted as the spatial component of a four-vector $Y$, which transforms linearly under the action of the boost generators.

To see this, notice that the time-time component $g_{00}$ of the spacetime metric $g_{\mu\nu}=e_\mu^I\eta_{IJ}e_\nu^J$, can be expressed in terms of $\chi^i=-e^i_0/e^0_0$ as
\be\nonumber
g_{00}=\left(e_0^0\right)^2\left(1-\chi^2\right),
\ee
which implies that $\chi^2\leq 1$. As a consequence,  $\chi$ cannot be directly interpreted as the spatial component $Y\in\mathbb R^3$ of a four-vector $V=(T,Y)$, with $T \in\mathbb R$, because the spatial norm $Y^2$ of any four-vector is not Lorentz-invariant. However, if $V$ is a timelike vector of unit norm, then $T^2-Y^2=1$, and $\chi$ can be interpreted as its velocity $Y/T$. In that case, the inequality $\chi^2\leq 1$ makes sense because $1-\chi^2=1/T^2\leq 1$ is a Lorentz-invariant inequality. For this reason, we perform the change of variables
\be\nonumber
\chi\longrightarrow Y\equiv T\chi,\qquad\qquad
\text{where}\qquad\qquad
T=\sqrt{1+Y^2}=\f{1}{\sqrt{1-\chi^2}}.
\ee
With these variables, the action of the boosts and the rotations simplifies, and is given by
\ba\nonumber
\big\{\mathcal{B}(u),T\big\}=Y\cdot u,\qquad
\big\{\mathcal{B}(u),Y\big\}=Tu,\qquad
\big\{\mathcal{R}(u),T\big\}=0,\qquad
\big\{\mathcal{R}(u),Y\big\}=Y\wedge v.
\ea
Therefore, we can build with the scalar $T$ and the three-vector $Y$, a timelike four-vector $V=(T,Y)$ of unit norm, on which the Lorentz group acts linearly. In matricial notations, we have
\be\nonumber
(T,Y)
\begin{pmatrix}
 0 & u\tra \\
 u & \vb
\end{pmatrix}
=(Y\cdot u,Tu+Y\wedge v).
\ee
Here, we have associated to any vector $v=(v_1,v_2,v_3)\in\mathbb R^3$ the three-dimensional antisymmetric rotational matrix $\vb$ (see appendix \ref{lorentz algebra}). As a consequence, the vector $\chi$ is naturally associated to a boost $B(\chi)$. This latter is uniquely defined by the relation
\be\nonumber
(T,Y)=B(\chi)(1,0),
\ee
and given in the four-dimensional vectorial representation by the block matrix
\be\nonumber
B(\chi)=
\begin{pmatrix}
 T & Y\tra \\
 Y &~~~\displaystyle 1+\f{1}{T+1}(Y\cdot Y\tra)
\end{pmatrix}.
\ee
Notice that this boost is clearly the same as the one we have introduced formally in (\ref{discussion on B}), since its inverse sends $\chi$ to the null vector, and therefore maps the general cotetrad $e$ to the time-gauge cotetrad $\tilde{e}$. This boost $B(\chi)$ will play an important role in the rest of this work.

Before studying the connection variable $A$ in detail, let us finish the study of the variable $\chi$ with a discussion concerning its conjugate momentum $\zeta$. We have found with geometrical arguments that the variable $Y$ is more appropriate than the original variable $\chi$, but the problem is that $\zeta$ is not conjugated to $Y$ in the sense of the Poisson bracket. Therefore, we would like to find a new variable $Z$ which is conjugated to $Y$. In other words, we are looking for a canonical transformation which maps the initial pair of variables $(\chi,\zeta)$, to a new pair of variables $(Y,Z)$ with clear geometrical interpretation. As one can see from the expression (\ref{gauss}) of the boost and rotation generators, the natural candidate for $Z$ turns out to be
\be\nonumber
Z\equiv\frac{1}{T}(\zeta-(\zeta\cdot\chi)\chi), 
\ee
because the $\zeta$-dependent parts of $\mathcal{B}(u)$ and $\mathcal{R}(v)$ can, respectively, be written very simply in terms of $Z$ as
\be\nonumber
-\int d^3x\,Tu\cdot Z,\qquad\qquad\text{and}\qquad\qquad-\int d^3x\,Z\wedge Y.
\ee
One can see that $Z$ is canonically conjugated to $Y$, since their Poisson brackets are given by
\ba\nonumber
\big\{Y_i(x),Z^j(y)\big\}=\delta^j_i\delta^3(x-y),\qquad\qquad
\big\{Y_i(x),Y_j(y)\big\}=0,\qquad\qquad
\big\{Z^i(x),Z^j(y)\big\}=0.
\ea
If one can interpret $Y$ as the spatial component of a four-vector $V=(T,Y)$ according to the previous considerations, it is not the case for its conjugate momentum $Z$. Indeed, $Z$ transforms as a vector under the action of the rotations, but its transformation law under the boosts is much more complicated, and involves the variable $E$. For this reason, it is very difficult to find a geometrical interpretation for $Z$, as we did for $Y$.

\subsection{General form of the Lorentz connection $\mathcal{A}$}

\noindent Now we address the problem of finding a Lorentz-covariant connection for canonical gravity. More precisely, we give in this subsection the general form of a Lorentz connection which depends only on $A$ and $\chi$. Before doing so, we are going to recall the equations that a connection has to satisfy in order to have the property of Lorentz covariance.

A general Lorentz connection $\mathcal{A}$ possesses two components. In terms of the boost and rotational components $\bst\mathcal{A}$ and
$\rot\mathcal{A}$, it can be written as
\be\nonumber
\mathcal{A}=\bst\mathcal{A}^iP_i+\rot\mathcal{A}^iJ_i,
\ee
where the basis $(J_i,P_i)$ with $i\in\{1,2,3\}$ of the Lorenz algebra $\so(3,1)$ has been defined in appendix \ref{lorentz algebra}. Depending on whether we are considering the abstract Lie algebra element, the vectorial or the spinorial representation, $\bst\mathcal{A}$ and $\rot\mathcal{A}$ will refer to elements of $\so(3,1)$, vectors in $\mathbb R^3$, or two-dimensional complex and traceless matrices. Therefore, we choose the same notation for these three different interpretations, but the appropriate meaning will always be clear from the context.

With these notations, $\mathcal A$ is a Lorentz connection if it satisfies the condition 
\be\nonumber
\big\{\mathcal{B}(u)+\mathcal{R}(v),\mathcal{A}\big\}
\, = \, d u+\rot\mathcal{A}\wedge u+\bst\mathcal{A}\wedge v+dv+\rot\mathcal{A}\wedge v-\bst\mathcal{A}\wedge u.
\ee
Here we have identified the components of $\mathcal A$ with vectors in $\mathbb R^3$.
Therefore, we require the boost and rotation parts to transform respectively as
\begin{subequations}\label{transformation}
\ba
\big\{\mathcal{B}(u)+\mathcal{R}(v),\bst\mathcal{A}\big\}&=&du+\rot\mathcal{A}\wedge u+\bst\mathcal{A}\wedge v=du-\ub\rot\mathcal{A}-\vb\,\bst\mathcal{A},\\
\big\{\mathcal{B}(u)+\mathcal{R}(v),\rot\mathcal{A}\big\}&=&dv+\rot\mathcal{A}\wedge v-\bst\mathcal{A}\wedge u=dv-\vb\,\rot\mathcal{A}+\ub\bst\mathcal{A}.
\ea
\end{subequations}
Now, our problem  consists in finding the general solution of these covariant equations.
Each of the two components can be identified with one-forms taking values in $\mathbb R^3$. Therefore, our problem reduces 
to finding two $\mathbb R^3$-valued one-forms which depends only on $A$ and $\chi$. For the reasons that have been invoked in the previous 
subsection, we are going to look for a connection that depends on $A$ and $Y$ (instead of $\chi$).

We have two natural $\mathbb{R}^3$-valued one-forms at our disposal: the variable $A$ itself, and the exterior derivative $dY$ of the vector $Y$. As a consequence, the general solution for $\bst\mathcal{A}$ or $\rot\mathcal{A}$, generically denoted $\brA$, is necessarily of the form
\be\nonumber
\brA=MA+NdY,
\ee
where $M$ and $N$ are two $3\times 3$ matrices. The only matrices that we can construct from $Y$ and $A$, are $\Yb$ itself, and any power $\Yb^\alpha$, where $\alpha$ can be \textit{a priori} any real number. However, as one can see from the characteristic polynomial
\be\label{polynomial}
\Yb^3+Y^2\Yb=0,
\ee
where $Y^2=Y^iY_i$ is the square norm of the vector $Y$, the matrix $\underline{Y}$ is not invertible, and admits purely imaginary nontrivial eigenvalues. This shows that we cannot consider any arbitrary real power $\Yb^\alpha$. The exponent $\alpha$ must be nonnegative, and also an integer in order to have a connection with values in $\mathbb R^3$ and not in $\mathbb C^3$. In summary, because of the form of the characteristic polynomial (\ref{polynomial}), it turns out that the most general $\mathbb R^3$-valued one-form can be written as
\be\label{generalform}
\brA=\left(a_0+a_1\Yb+a_2\Yb^2\right)A+\left(b_0+b_1\Yb+b_2\Yb^2\right)dY,
\ee
where $a_i$ and $b_i$ are \textit{a priori} functions of the coordinates $Y_i$ only. Using the standard properties of the matrices $\underline{Y}$
and $\Yb^2$, one can write the previous expression for the components of the connection as
\be\label{general expression 2}
\brA=\tilde{a}_0A+a_1Y\wedge A+\tilde{b}_0dY+b_1Y\wedge dY+\big( a_2(A\cdot Y)+b_2(Y\cdot dY)\big)Y,
\ee
where $\tilde{a}_0=a_0-a_2Y^2$ and $\tilde{b}_0=b_0-b_2Y^2$. 

Now, we can look for a Lorentz-covariant connection $\mathcal{A}$ admitting a pure boost component $\bst\mathcal{A}$ and a pure rotational component $\rot\mathcal{A}$, both being of the form (\ref{generalform}).

\subsection{Covariance under the rotations}

\noindent The condition of covariance under the action of the rotations implies that $\brA$ has to satisfy
\be\label{rotation covariance}
 \big\{\mathcal{R}(v),\brA\big\}=Hdv-\vb\brA,
\ee
where the constant $H$ has to be $H=0$ for $\bst\mathcal{A}$, and $H=1$ for $\rot\mathcal{A}$. To show if this is possible, we compute the action of the generator of the rotations on the form (\ref{generalform}) of $\brA$.

One can immediately see that the coefficients $a_i$ and $b_i$ must be invariant under the action of the rotations, otherwise there would be no nontrivial solution to equation (\ref{rotation covariance}). As a consequence, we will impose from now that these coefficients are functions of the variable $T$ only. With this condition, and using the formulas given in appendix \ref{formulas}, the action of the rotations on $\brA$ turns out to be
\ba
\big\{\mathcal{R}(v),\brA\big\}&=&\left(a_1[\Yb,\vb]+a_2[\Yb^2,\vb]\right)A+\left(b_1[\Yb,\vb]+b_2[\Yb^2,\vb]\right)dY\nonumber\\
&&+\left(a_0+a_1\Yb+a_2\Yb^2\right)\left(-\frac{1}{\gamma}dv+\f{1}{T}\Yb dv-\vb A\right)\nonumber\\
&&+\left(b_0+b_1\Yb+b_2\Yb^2\right)\left(-\vb dY+\Yb dv\right).\nonumber
\ea
After some straightforward calculations, one ends up with the expression
\ba
\big\{\mathcal{R}(v),\brA\big\}&=&\left(-\frac{a_0}{\gamma}+\left(\frac{a_0}{T}-\frac{a_1}{\beta}-\frac{a_2}{T}Y^2 
+b_0-b_2 Y^2\right)\Yb+\left(\frac{a_1}{T}-\frac{a_2}{\gamma}+ b_1\right)\Yb^2\right)dv\nonumber\\
&&-\vb\left(a_0+a_1\Yb+a_2\Yb^2\right)A -\vb\left(b_0+b_1\Yb+b_2\Yb^2\right).\nonumber
\ea
We have used the characteristic equation  $\Yb^3+Y^2\Yb=0$ to get this result. 
From this, it is immediate to see that the covariance under rotations (\ref{rotation covariance}) holds if and only if
the coefficients entering in the definition of $\brA$ satisfy the relations
\be\label{conditions rotations 1}
-\f{a_1}{\gamma}+\f{a_0}{T}-\f{a_2}{T}Y^2+b_0-b_2Y^2=0,\qquad\text{and}\qquad\qquad
-\f{a_2}{\gamma}+\f{a_1}{T}+b_1=0,
\ee
together with 
\be\label{conditions rotations 2}
-\frac{a_0}{\gamma}=H.
\ee
Therefore, in order to have the right transformation property, i.e. equation (\ref{transformation}), we need to set $a_0=-\gamma$ for the component $\rot\mathcal{A}$, and $a_0=0$ for the component $\bst\mathcal{A}$ (if $\gamma\neq\infty$).
As a consequence, the parameters $a_i$ are completely fixed in terms of the parameters 
$b_i$. We see that the requirement of covariance under the rotations fixes six parameters out of the 12 that were introduced initially in the definition of $\brA$. 

\subsection{Covariance under the boosts}

\noindent Studying the covariance under the action of the boost generator is much more involved than studying the covariance
under the rotations. The reason for this is simple: the action of the boost generator ${\mathcal{B}(u)}$ on the phase space variables
is complicated (see appendix \ref{formulas}). It is nonetheless possible to solve this problem, and we are going to show that the requirement of the
covariance under the boosts determines uniquely the Lorentz connection $\mathcal{A}$. As the calculations are long,
we will proceed in two steps for the sake of clarity. First, we will show that the rotational and boost components of the
connection each depend on two real parameters. Then, we will show that the requirement of Lorentz covariance fixes
uniquely these parameters to values that depend on the Immirzi-Barbero parameter $\gamma$ only.

We start with the general expression (\ref{general expression 2}) of the connection $\mathcal A$.
The condition of covariance of the components $\brA$ under the boosts is
\be\nonumber
\big\{\mathcal{B}(u),\brA\big\}=(1-H)du+(2H-1)\ub\rbA.
\ee
Here, we have used the same notation as in equation (\ref{rotation covariance}), where $H=0$ (respectively $H=1$) when $\mathcal{B}(u)$
acts on the boost (respectively rotational) component of $\mathcal A$. Before going into further details, it is interesting to note that, contrary to the previous case, the requirement of covariance under the boosts mixes the boost and rotational components of the Lorentz connection. 

\subsubsection*{\textbf{First step: the two-parameter family}}

\noindent Using the results given in appendix \ref{formulas}, it is long but straightforward to show that 
\be\label{good bad}
\big\{\mathcal{B}(u),\brA\big\}=\left(\tilde{a}_0+\tilde{b}_0-\frac{Y^2}{\gamma T}a_1\right)du+\mathcal{Y}+\mathcal{Z}
\ee
where $\mathcal{Y}$ is a one-form with a ``good shape'':
\ba
\mathcal{Y}&=&a_1Tu\wedge A+b_1Tu\wedge dY-\left(a_1\frac{A\cdot Y}{T}+b_1 dT\right)u\wedge Y\nonumber\\ 
&&+(Y\cdot u)\left(\tilde{a}_0'-\frac{\tilde{a}_0}{T}\right)A+a_2T(A\cdot u)Y+(Y\cdot u)\tilde{b}'_0 dY+b_2T(dY\cdot u)Y,\label{good part}
\ea
and $\mathcal{Z}$ is a one-form with a ``bad shape'':
\ba
\mathcal{Z}&=&(Y\cdot u)\left(a'_1-\frac{a_1}{T}\right)Y\wedge A+(Y\cdot u)b'_1Y\wedge dY+\left(a_1+\frac{\tilde{a}_0}{\gamma T}
+b_1T\right)Y\wedge du\nonumber\\
&&+\left(a_2'(Y\cdot u)(A\cdot Y)+(Y\cdot u)\big(b'_2(Y \cdot dY)+b_2 dT\big)+\left(\frac{a_1}{\gamma T}+a_2+b_2T\right)(du\cdot Y) \right)Y\nonumber\\
&&+\left((A\cdot Y)\left(\frac{\tilde{a}_0}{T}+a_2T\right)+\tilde{b}_0dT+b_2T(Y\cdot dY)\right)u.\nonumber
\ea
The ``good shape'' part of (\ref{good bad}) regroups all the terms appearing in the expansion of $\pm\ub\brA$.
On the contrary, the ``bad shape'' part regroups all the terms that cannot arise from the expansion of $\pm \ub  \brA $.
As a consequence, for the connection to satisfy the property of covariance under the boosts, the term $\mathcal{Z}$ has to vanish.
Thus, each independent term in $\mathcal{Z}$ has to vanish. Because of the fact that $Y\cdot dY=TdT$ (which is a consequence of
the ``mass-shell'' relation $T^2-Y^2=1$), there are 8 independent terms in $\mathcal{Z}$, and therefore we end up with the 8 conditions
\be\nonumber
\begin{array}{rrrr}
\displaystyle a'_1-\frac{a_1}{T}=0, & 
\qquad\qquad a'_2=0, & 
\qquad\qquad Tb'_2+b_2=0, & 
\qquad\qquad\displaystyle\frac{a_1}{\gamma T}+a_2+Tb_2=0, \\
\displaystyle\frac{\tilde{a}_0}{T}+a_2T=0, & 
b'_1=0, & 
\tilde{b}_0+T^2b_2=0, & 
\displaystyle a_1+\frac{\tilde{a}_0}{\gamma T}+b_1T=0.
\end{array}
\ee
The general solution to this system of linear equations is a two-dimensional vector space. Given two real parameters $x$ and $y$, the solution is given by
\begin{subequations}\label{partial solution}
\ba
\tilde{a}_0=-xT^2, \qquad\qquad & 
a_1=-\gamma(x+y)T, \qquad\qquad & 
a_2=x, \\
\tilde{b}_0=-yT, \qquad\qquad &
\displaystyle b_1=\left(\frac{1}{\gamma}+\gamma\right)x+\gamma y, \qquad\qquad & 
\displaystyle b_2=\frac{y}{T}.
\ea
\end{subequations}
As a consequence, each of the two components $\brA$ of the Lorentz connection depend at this stage only 
on a pair of real parameters $(x,y)$. The explicit expression of $\brA$ in terms of $x$ and $y$ is given by
\ba
\brA(x,y)&=&-xT^2A-yTdY-\gamma(x+y)TY\wedge A\nonumber \\
&&+x(A\cdot Y)Y+\left[\left(\frac{1}{\gamma}+\gamma\right)x+\gamma y\right]Y\wedge dY+ydTY. \label{brA vector}
\ea
In matricial notations, this expression becomes
\be
\brA(x,y)=\left(-x-\gamma(x+y)\Yb+x\Yb^2\right)A
+\left(-\frac{y}{T}+\left[\left(\frac{1}{\gamma}+\gamma\right)x+\gamma y\right]\Yb+\frac{y}{T}\Yb^2\right)dY.\label{brA matrix}
\ee
At this point, it is important to underline that imposing such an expression for the boost and rotational components of the connection $\mathcal{A}$,
is only a necessary condition (and not a sufficient one) for $\mathcal{A}$ to transform in a covariant way under the action of the boosts.

For the sake of clarity, we will use from now on the notation $\mathcal{C}(x,y)$ for the general expression of $\brA(x,y)$ given in (\ref{brA vector}) and (\ref{brA matrix}). 

\subsubsection*{\textbf{Second step: fixing uniquely the parameters}}

\noindent The issue of finding a covariant connection under the action of the boosts, now reduces to that of finding two pairs, $(x_1,y_1)$ and $(x_2,y_2)$, defining $\bst\mathcal{A}=\mathcal{C}(x_1,y_1)$ and $\rot\mathcal{A}=\mathcal{C}(x_2,y_2)$ in such a way that
\be\label{equations of covariance involving C}
\big\{\mathcal{B}(u),\mathcal{C}(x_1,y_1)\big\}=du-\ub\mathcal{C}(x_2,y_2),\qquad\qquad\text{and}\qquad\qquad
\big\{\mathcal{B}(u),\mathcal{C}(x_2,y_2)\big\}=\ub\mathcal{C}(x_1,y_1).
\ee
To solve this problem, we need to know how $\mathcal{C}(x,y)$ transforms under the action of the boosts. Using equation (\ref{good bad}), and the solutions (\ref{partial solution}) into the expression (\ref{good part}) for $\mathcal{Y}$, one obtains immediately that
\be\nonumber
\big\{\mathcal{B}(u),\mathcal{C}(x,y)\big\}=-(x+y)du+\ub\widetilde{\mathcal{C}}(x,y),
\ee
where the $\mathbb{R}^3$-valued one form $\widetilde{\mathcal{C}}(x,y)$ is given by
\ba
\widetilde{\mathcal{C}}(x,y)(x,y)&=&-\gamma(x+y)T^2A+\left[\left(\frac{1}{\gamma}+\gamma\right)x+\gamma y\right]TdY+xTY\wedge A \nonumber \\
&&+\gamma(x+y)(A\cdot Y)Y+yY\wedge dY-\left[\left(\frac{1}{\gamma}+\gamma\right)x+\gamma y\right]dTY.\nonumber
\ea
With the matricial notation, $\widetilde{\cal C}(x,y)$ reads
\ba
\widetilde{\mathcal{C}}(x,y)(x,y)&=&\left(-\gamma(x+y)+xT\Yb+\gamma(x+y)\Yb^2\right)A \nonumber \\
&&\left(\frac{1}{T}\left[\left(\frac{1}{\gamma}+\gamma\right)x+\gamma y\right]+y\Yb-\frac{1}{T}\left[\left(\frac{1}{\gamma}+\gamma\right)x+\gamma y\right]\Yb^2\right)dY.\nonumber
\ea

As a consequence, the equations (\ref{equations of covariance involving C}) of covariance under the boosts imply that
\ba
-(x_1+y_1)du+\ub\widetilde{\mathcal{C}}(x_1,y_1)&=&du-\ub{\mathcal{C}}(x_2,y_2), \nonumber\\
-(x_2+y_2)du+\ub\widetilde{\mathcal{C}}(x_2,y_2)&=&\ub{\mathcal{C}}(x_1,y_1).\nonumber
\ea
Therefore, the two pairs of parameters $(x_1,y_1)$ and $(x_2,y_2)$ are determined by the system of equations
\be\nonumber
x_1+y_1=-1,\qquad x_2+y_2=0,\qquad\widetilde{\mathcal{C}}(x_1,y_1)=-\mathcal{C}(x_2,y_2),\qquad 
\widetilde{\mathcal{C}}(x_2,y_2)=\mathcal{C}(x_1,y_1).
\ee
At first sight, there are many more relations than unknowns in this system, and therefore the hope of finding even one
single solution is very low. However, very fortunately, it turns out that the system admits a unique solution, which is simply given by
\be\nonumber
x_1=0,\qquad\qquad y_1=-1,\qquad\qquad x_2=\gamma,\qquad\qquad y_2=-\gamma.
\ee
Finally, using this solution, we find that the boost and rotational components are given by
\ba
\bst\mathcal{A}&=&\gamma\,T\,\Yb A+\f{1}{T}\left(1-\gamma\,T\,\Yb-\Yb^2\right)dY,\nonumber\\
\rot\mathcal{A}&=&-\gamma\left(1-\Yb^2\right)A+\f{\gamma}{T}\left(1+\f{T}{\gamma}\,\Yb-\Yb^2\right)dY.\nonumber
\ea
The Lorentz-covariant connection  can be expressed in a simpler form if we replace the variable $Y$ by its expression in terms of the original vector $\chi$. Indeed, a simple calculation leads to
\be
\rot\mathcal{A}=\f{1}{1-\chi^2}\left(\chi\wedge d\chi+\Omega_0\right),\qquad\qquad
\bst\mathcal{A}=\f{1}{1-\chi^2}\left(d\chi+\Omega_0\wedge\chi\right),\label{uc}
\ee
with
\be\nonumber
\Omega_0=\gamma\left(d\chi-A+(A\cdot\chi)\chi\right).
\ee

It is remarkable that this unique connection transforms also correctly under the action of the rotations. Indeed, it is easy to show that the coefficients $a_i$ and $b_i$, determined implicitly above, satisfy the conditions (\ref{conditions rotations 1}) and (\ref{conditions rotations 2}). As a consequence, we end up with a unique Lorentz-covariant connection. In that sense, we claim that there is a unique commutative Lorentz-covariant connection in first-order general relativity.  Obviously, this connection reduces to the standard Ashtekar-Barbero connection when the time-gauge is imposed.

\subsection{Action of spatial diffeomorphisms}

\noindent In the last subsection, we found the unique Lorentz-algebra valued one-form $\mathcal{A}$ which transforms correctly under the action of the boost and rotation generators. However, this property of covariance is not sufficient to conclude that $\mathcal{A}$ is indeed a connection. The fact that $\mathcal{A}$ is a one-form is nonetheless a strong indication that $\mathcal{A}$ is
a connection. But, to prove it explicitly, it is necessary to show that $\mathcal{A}$ transforms correctly under the action of the constraints that generate spatial diffeomorphisms.

For that purpose, we have to exhibit out of the full set of first-class constraints of the theory, the constraint $\widetilde{\mathcal{H}}_a$ that generates the spatial diffeomorphisms. Then, we have to show that the Poisson action on $\mathcal{A}$ of the constraint smeared with a vector $N^a$, gives the Lie derivative of $\mathcal{A}$ along this vector. This is exactly what we are going to do here.

The smeared vector constraint is given by (\ref{vector constraint}) \cite{barros}. It is quite complicated and does not generate spatial diffeomorphisms. As usual, the generator of spatial diffeomorphisms turns out to be given by a linear combination of the vector and the Gauss constraints (\ref{gauss}). In the present context, it is given by the following combination of $\mathcal{H}_a$ and the generators $\mathcal{B}$ and $\mathcal{R}$ of the Lorentz algebra:
\ba\nonumber
\widetilde{\mathcal{H}}_a(N^a)&=&\mathcal{H}_a(N^a)-\f{\gamma}{1+\gamma^2}\Big(\gamma\mathcal{B}(N^aA_a)-\mathcal{R}(N^aA_a)\Big)\nonumber\\
&=&\int d^3x\,N^a\left(E^b\cdot\partial_aA_b+\zeta\cdot\partial_a\chi-E^b\cdot\partial_bA_a-A_a\cdot\partial_bE^b\right).\nonumber
\ea
Indeed, $\widetilde{\mathcal{H}}_a(N^a)$ satisfies the algebra of spatial diffeomorphisms, and its action on the variables
$A$ and $\chi$,
\ba
\big\{\widetilde{\mathcal{H}}_a(N^a),A_b\big\}&=&-N^a\partial_aA_b-A_a\partial_bN^a=-\pounds_{N^a}A_b,\nonumber\\
\big\{\widetilde{\mathcal{H}}_a(N^a),\chi\big\}&=&-N^a\partial_a\chi=-\pounds_{N^a}\chi,\nonumber
\ea
is clearly the Lie derivative along the vector field $N^a$ of a one-form and a scalar. We see that the phase space variables $A$ and $\chi$ can really be geometrically interpreted as a one-form and a scalar. It is therefore immediate to conclude that the connection $\mathcal{A}$ will also transform properly under the action of diffeomorphisms of the spatial hypersurface, because it depends only on $A$ and $\chi$.

Before ending this discussion, let us study how the momentum variables $E$ and $\zeta$ transform under the action of $\widetilde{\mathcal{H}}_a(N^a)$. A straightforward calculation leads to
\ba
\big\{\widetilde{\mathcal{H}}_a(N^a),E^b\big\}&=&-E^b\partial_aN^a-N^a\partial_aE^b+E^a\partial_aN^b=-\pounds_{N^a}E^b,\nonumber\\
\big\{\widetilde{\mathcal{H}}_a(N^a),\zeta\big\}&=&-\partial_a\left(N^a\zeta\right).\nonumber
\ea
As a result, $E$ transforms as expected as a vector field of density one. On the other hand, it is clear that $\zeta$ does not transform as a scalar, it is a density. It is nonetheless easy to see that the ``undensitized" quantity $\zeta/\sqrt{\zeta^2}$ does transform as a scalar. 

\section{Reduction to an $\SU(2)$ connection}
\label{section su(2)}

\noindent Having a Lorentz connection at our disposal, it is quite natural to wonder if we can act on it with a boost, to send it back to a pure $\su(2)$ connection. In view of the argument given in the introduction of section \ref{covariant connection}, there should be a relation between the Ashtekar-Barbero connection in the time-gauge, and the covariant connection we have just constructed. To make this relation concrete, we would like to ask the question whether it is possible to find a boost $B(n,\theta)$ that acts on the connection $\mathcal{A}$, and gives a connection $\tilde{\mathcal{A}}$ with vanishing boost component. The answer turns out to be affirmative, as we are going to show in this section.

Let us first introduce some notations that are more suited for the calculations. Since $\chi$ is a velocity ($\chi^2<1$), it is possible to write it as $\chi=n\,\htan$, where $\alpha$ is a real parameter, and $n$ is a vector in $\mathbb{S}^3$. Then, it is easy to see that
\be\nonumber
\f{1}{1-\chi^2}=\text{ch}^2\alpha,\qquad\qquad
d\chi=\f{d\alpha}{\text{ch}^2\alpha}n+dn\,\text{th}\,\alpha,\qquad\qquad
\chi\wedge d\chi=(n\wedge dn)\,\text{th}^2\,\alpha.
\ee
With these notations, the boost and rotational components (\ref{uc}) of the connection $\mathcal{A}$ take the form
\ba\nonumber
\bst\mathcal{A}&=&(d\alpha n+dn\,\text{ch}\,\alpha\,\text{sh}\,\alpha)+(\tilde{\Omega}_0\wedge n)\,\text{th}\,\alpha,\nonumber\\
\rot\mathcal{A}&=&(n\wedge dn)\,\text{sh}^2\alpha+\tilde{\Omega}_0,\nonumber
\ea
where
\be\nonumber
\tilde{\Omega}_0=\gamma\left(d\alpha n+dn\,\text{sh}\,\alpha\,\text{ch}\,\alpha-A\,\text{ch}^2\alpha+(A\cdot n)n\,\text{sh}^2\alpha\right).
\ee

With this form of the connection, and the action of the boost given in appendix \ref{boost action}, one can show that the boost $B(n,-\alpha)$ is precisely the one sending the boost component $\bst\mathcal{A}$ of the Lorentz-covariant connection to
\be\nonumber
\bst\tilde{\mathcal{A}}\equiv \bst\,\,(B(n,-\alpha)\triangleright\mathcal{A})=0.
\ee
The notation $B(n,\theta)$ holds for the boost with ``angle'' $\theta$ and direction $n$, and the action of any boost $B$ and
a Lorentz connection $\cal A$ is given as usual by 
\be\nonumber
B\triangleright\mathcal{A}=B^{-1}dB+B^{-1}\mathcal{A}B.
\ee
This boost has a very natural physical interpretation as the boost sending the four-vector $(T,Y)$ to $(1,0)$. In other words, it maps the variable $\chi$ to zero and, in a sense, maps the non time-gauge formalism to the time-gauge one. Under the action of this boost, the rotational component $\rot\mathcal{A}$ of the Lorentz-covariant connection transforms as
\ba\nonumber
\rot\tilde{\mathcal{A}}\equiv \rot \,\,(B(n,-\alpha)\triangleright\mathcal{A})&=&
\gamma\left(nd\alpha+dn\,\text{sh}\,\alpha-A\,\text{ch}\,\alpha+(\text{ch}\,\alpha-1)(A\cdot n)n\right)+(\text{ch}\,\alpha-1)n\wedge dn\nonumber\\
&=&-\gamma\left(A\,\text{ch}\,\alpha+(1-\text{ch}\,\alpha)(A\cdot n)n\right)+f(\chi),\nonumber
\ea
where the one-form $f(\chi)$ is the part of the connection that does not depend on $A$ (and therefore depends only on $\chi$). This expression will play a capital role in the quantum theory. It will be convenient to use the form
\be\label{su2 part}
\rot\tilde{\mathcal{A}}=-\gamma\mu A+f(\chi),
\ee
where the matrix $\mu$ is given by
\be\nonumber
\mu=\openone\,\text{ch}\,\alpha+N(1-\text{ch}\,\alpha),
\ee
with $N_{ij}=n_in_j$, and $\openone$ the identity matrix.

\section{Covariant  momenta}
\label{moments}

\noindent At this point, we have at our disposal a Lorentz-covariant connection depending upon the configuration variables $A$ and $\chi$, and we have not said anything about the momentum variable $E$. It is now therefore natural to use the construction of section \ref{covariant connection} to build a function of $E$ and $\chi$ that transforms in a covariant way under the Lorentz generators. We will then see that this variable is related to the surface operator in a very nice way, and derive the implications for the quantum theory.

\subsection{General form of the momenta and the two-parameter family of solutions}

\noindent Using the terminology that we have introduced to construct the Lorentz connection, we can decompose the ``momentum'' variable that we try to build into its boost and rotational parts as
\be\nonumber
\mathcal{E}=\bstE^iP_i+\rotE^iJ_i.
\ee
With this notation, the condition of covariance under the action of the Lorentz generators takes the form
\be\nonumber
\big\{\mathcal{B}(u)+\mathcal{R}(v),\mathcal{E}\big\}=\rotE\wedge u+\bstE\wedge v+\rotE\wedge v-\bstE\wedge u,
\ee
where we have identified the components of $\mathcal{E}$ with vectors in $\mathbb{R}^3$. 
Therefore, the boost and rotation parts transform respectively as
\begin{subequations}\label{transformation de E}
\ba
\big\{\mathcal{B}(u)+\mathcal{R}(v),\bstE\big\}&=&\rotE\wedge u+\bstE\wedge v,\\
\big\{\mathcal{B}(u)+\mathcal{R}(v),\rotE\big\}&=&\rotE\wedge v-\bstE\wedge u.
\ea
\end{subequations}
The ansatz that we make for the general form of the solution is simply
\begin{subequations}\label{generalE}
\ba
\bstE&=&c_1E+c_2Y\wedge E,\\
\rotE&=&d_1E+d_2Y\wedge E.
\ea
\end{subequations}
We see immediately that the variable $\zeta$ does not appear in the expression of $\cal{E}$. The reason is that $\zeta$ transforms in a very complicated way under the boosts and the rotations, and has no obvious physical interpretation. Moreover, the ansatz that we choose will lead to a very simple and nice solution for the Lorentz-covariant momenta variables. At this point, it is important to notice that the variables $E$ and $\chi$ (or $Y$ equivalently) already have the right transformation behavior under the action of the generator of the rotations (see appendix \ref{formulas}). Therefore, it is sufficient to study the invariance of the expression (\ref{generalE}) under the action of the boosts.

Plugging the expressions (\ref{generalE}) into (\ref{transformation de E}) immediately leads to the following action of the boost generator:
\ba
\big\{\mathcal{B}(u),\bstE\big\}&=&\rotE\wedge u=d_1E\wedge u+d_2(Y\wedge E)\wedge u=d_1E\wedge u+d_2E(Y\cdot u)-d_2Y(E\cdot u),\nonumber\\
\big\{\mathcal{B}(u),\rotE\big\}&=&-\bstE\wedge u=-c_1E\wedge u-c_2(Y\wedge E)\wedge u=-c_1E\wedge u-c_2E(Y\cdot u)+c_2Y(E\cdot u).\nonumber
\ea
Now, this expression is to be compared with the one obtained by acting explicitly with the boost generator on the components (\ref{generalE}). Using the formulas given in appendix \ref{formulas}, one finds that
\ba
\big\{\mathcal{B}(u),\bstE\big\}&=&(Y\cdot u)\left(c_1'E+c_2'Y\wedge E\right)+\f{c_1}{T}\big[(Y\cdot u)E-(E\cdot u)Y\big]+c_2\left(Tu\wedge E+\f{1}{T}(Y\cdot u)Y\wedge E\right)\nonumber\\
&=&(Y\cdot u)\left(c_1'+\f{c_1}{T}\right)E+(Y\cdot u)\left(c_2'+\f{c_2}{T}\right)Y\wedge E+c_2Tu\wedge E-\f{c_1}{T}(E\cdot u)Y,\nonumber
\ea
and the same expression for $\rotE$ if we replace the $c$ coefficients by the $d$ ones. It is then immediate to see that in order for the components $\bstE$ and $\rotE$ to have the right transformation behavior under the action of the boosts, the various $c$ and $d$ coefficients have to satisfy
\be\nonumber
\begin{array}{rrrr}
\displaystyle c'_1+\frac{c_1}{T}=d_2, & 
\qquad\qquad\displaystyle c'_2+\frac{c_2}{T}=0, & 
\qquad\qquad c_2T=-d_1, & 
\qquad\qquad\displaystyle -\frac{c_1}{T}=-d_2, \\\\
\displaystyle d'_1+\frac{d_1}{T}=-c_2, & 
\qquad\qquad\displaystyle d'_2+\frac{d_2}{T}=0, & 
\qquad\qquad d_2T=c_1, & 
\qquad\qquad\displaystyle -\frac{d_1}{T}=c_2.
\end{array}
\ee
The solution to this system of equations is given in terms of two real parameters $(\alpha_1,\alpha_2)\in\mathbb{R}^2$ as
\be\nonumber
c_1=\alpha_1,\qquad\qquad c_2=\f{\alpha_2}{T},\qquad\qquad d_1=-\alpha_2,\qquad\qquad d_2=\f{\alpha_1}{T}.
\ee
The covariant boost and rotational components are therefore given by
\ba
\bstE&=&\alpha_1E+\f{\alpha_2}{T}Y\wedge E,\nonumber\\
\rotE&=&-\alpha_2E+\f{\alpha_1}{T}Y\wedge E.\nonumber
\ea
It is in fact quite natural to obtain a two-parameter family of solutions since the action (\ref{transformation de E}) is linear, and because given a solution $\mathcal{E}$, its dual $\star\mathcal{E}$ is also a solution. Therefore, up to Hodge transformations and linear transformations, the momentum variable $\mathcal{E}$ is unique, as the Lorentz-covariant connection $\mathcal{A}$.

\subsection{Interpretation}

\noindent The components of the two-parameter family of conjugate momenta can be written with the matricial notation as
\ba
\bstE&=&\left(\alpha_1\openone+\f{\alpha_2}{T}\Yb\right)E,\nonumber\\
\rotE&=&\left(-\alpha_2\openone+\f{\alpha_1}{T}\Yb\right)E.\nonumber
\ea
Now, given the $\mathfrak{sl}(2,\mathbb{C})$ gauge-invariant element $\mathcal{E}=\bstE^iP_i+\rotE^iJ_i$, we can construct the two quadratic Casimir functions
\ba
C_1&=&\rotE\cdot\bstE=-\alpha_1\alpha_2E\cdot\left(\openone +\f{1}{T^2}\Yb^2\right)E,\nonumber\\
C_2&=&\rotE^2-\bstE^2=\left(\alpha_2^2-\alpha_1^2\right)E\cdot\left(\openone +\f{1}{T^2}\Yb^2\right)E.\nonumber
\ea
We are naturally interested in these Casimir functions because they offer simple Lorentz-invariant functions and therefore they are kinematical observables of the theory. The question we address now is the physical meaning of these functions. From the equations above, we see that the Casimir functions turn out to be proportional, and an explicit computation shows that
\be\label{EetaE}
E\cdot\left(1+\f{1}{T^2}\Yb^2\right)E=E_i\eta^{ij}E_j,
\ee
where the matrix $\eta$ is given by
\be\label{metric}
\eta^{ij}=(1-\chi^2)\delta^{ij}+\chi^i\chi^j.
\ee
It is remarkable that the right-hand side of (\ref{EetaE}) is exactly the term appearing in the definition of the spatial metric $q_{ab}$. Indeed, we have the relation
\be\label{areaclassical}
\det(q_{ab})q^{ab}=E^a_i\eta^{ij}E^b_j,
\ee
where $q^{ab}$ is the inverse of $q_{ab}$, and the quantity so defined is therefore an invariant of $\SL(2,\mathbb{C})$. As a consequence, the Casimir functions that we have introduced enter the definition of the area of any surface $S$ according to the relation
\be\nonumber
\text{Area}(S)=\int_S\sqrt{\det(h_{ab})}=\int_S d^2 x \, \sqrt{E_i^a \eta^{ij}E_j^{b}n_an_b},
\ee
where $h_{ab}$ is the intrinsic two-metric induced by $q_{ab}$ on the surface $S$, whose unit normal vector is $n_a$. In that sense, the (square root of) Casimir functions associated to the momenta variables are proportional to the local area of a spatial surface $S$. We are going to exploit this property in the next section.

\section{Quantum theory and the area operator}

\noindent In usual loop quantum gravity, the fundamental variables are taken to be the holonomy $H_\ell(A)\in\SU(2)$ of the $\su(2)$ Ashtekar-Barbero connection $A^i_a$ along a path $\ell$, and the fluxes of the electric field $E^i_a$. These variables form a Poisson algebra, which is represented at the quantum level by the well-known holonomy-flux algebra. Among the most important results related to the study of this quantum algebra \cite{al,rovelli,thiemann}, are the construction  of the kinematical Hilbert space, the proof of the uniqueness of the associated representation \cite{lost}, and the calculation of the spectrum of geometric operators \cite{ashtekar-lewandowski,rovelli-smolin}. Without entering into the details, let us recall that kinematical states are described in terms of cylindrical functions, and that the Plancherel theory for the compact group $\SU(2)$ enables one to construct a basis of the kinematical Hilbert space in terms of spin-network states. Such a state is characterized by a topological graph $\Gamma$ (embedded in the spatial manifold), whose links $\ell$ are labeled by spin-$j$ representations of $\SU(2)$, and whose nodes $n$ are labeled by $\SU(2)$ intertwiners. Furthermore, spin-network states diagonalize the area operators of any surface $S$, and each link $\ell$ of the underlying graph contributes to the area of the surface $S$ with an elementary area given by
\be\label{area contribution}
a_\ell(j)=\gamma\ell_\text{Pl}^2 \sqrt{j(j+1)},
\ee
where $j$ is the representation carried the link $\ell$, and $\ell_\text{Pl}$ is the Planck length.

The  quantization of the Lorentz-covariant theory that we have constructed will follow the same lines. Contrary to what happens in the formalism developed by Alexandrov, which is based on a rather complicated Dirac bracket, here the nonphysical phase space (\ref{phase space}) is very easy to quantize. We make the following choice of polarization. The commuting variables $A$ and $\chi$ are chosen to be the configuration variables, whereas $E$ and $\zeta$ are momentum variables acting as derivative operators on any function $\Psi(\mathcal{A},\chi)$ according to the usual quantization rule
\be
\hat{E}^a_i\Psi(\mathcal{A},\chi)=-i\hbar\f{\delta}{\delta A^i_a}\Psi(\mathcal{A},\chi),\qquad\qquad
\hat{\zeta}_i\Psi(\mathcal{A},\chi)=-i\hbar\f{\delta}{\delta \chi^i}\Psi(\mathcal{A},\chi).
\ee
Following $\SU(2)$ loop quantum gravity, we make the assumption that the states $\Psi(\mathcal{A},\chi)$ are polymerlike functions with support on one-dimensional spatial links only. Let us make this statement more precise. First of all,  it is clear that we need a connection to do so and, naturally, we will use the unique Lorentz-covariant connection $\mathcal{A}$ that we have constructed. With this connection at hand, it is natural to consider the holonomy
\be\nonumber
 H_\ell(\mathcal{A})=\mathcal{P}\exp\int_\ell\mathcal{A}=\mathcal{P}\exp\int_\ell\left(\bst\mathcal{A}^iP_i+\rot\mathcal{A}^iJ_i\right)
\ee
along a link $\ell$. With this $\SL(2,\mathbb{C})$ holonomy, we consider spin-network states over the Lorentz group. As a consequence, our starting point to construct the covariant kinematical Hilbert space is the space of cylindrical functions $\Psi_\Gamma( H_\ell(\mathcal{A}),\chi)$ characterized by a graph $\Gamma$ in the spatial manifold. We have not discussed the dependence of the states on the variable $\chi$ because we are going to see that it is in fact not relevant. Nevertheless, let us point out that, as it is the case with the projected spin-networks \cite{alexandrov4}, the dependence of the cylindrical function $\Psi_\Gamma$  in $\chi$ is encoded only at the vertices of the graph $\Gamma$. Indeed, since $\chi$ is a scalar, it is not canonically associated to one-dimensional links. The spatial derivative $\partial_\mu\chi$ is naturally associated to one-dimensional links, but its integration along any link reduces to a contribution at the end points only. This does not constitute a proof, but gives a simple argument to understand why the dependence on $\chi$ appears only at the vertices of $\Gamma$.

Lorentz gauge-invariant functions can be obtained from the previous spin-network states by requiring gauge invariance at the vertices of the graph $\Gamma$. The vector space of Lorentz gauge-invariant states is well-defined, but it is not possible to consistently define a Hilbert space structure over the whole space of cylindrical functions due to the noncompactness of the gauge group \cite{freidel-livine}. Fortunately, we have seen in section \ref{section su(2)} that it is possible to exhibit a boost $B(n,-\alpha)$ which sends the connection $\mathcal{A}$ to a connection $\tilde{\mathcal{A}}$ with vanishing boost component. Furthermore, the action of this boost maps $\chi$ to zero, and therefore eliminates the explicit dependence of the function $\Psi$ in $\chi$. Note however that $\chi$ appears implicitly in the expression of $\mathcal{A}$. As a consequence, any Lorentz-covariant kinematical state $\Psi_\Gamma(\cal{A},\chi)$ is canonically associated with a usual $\SU(2)$ kinematical state 
\be\label{holonomy}
\widetilde{\Psi}_\Gamma(\tilde{\mathcal{A}})=\Psi_\Gamma\left(B^{-1}_{s(\ell)}H_\ell(\mathcal{A})B_{t(\ell)},B^{-1}\chi\right)
=\Psi_\Gamma(H_\ell(\cal{A}),\chi).
\ee
As a conclusion, we can construct a Hilbert space structure on the space of Lorentz gauge-invariant states. It is isomorphic to the $\SU(2)$ kinematical Hilbert space, and a basis is given by the usual $\SU(2)$ spin-network states. Before going further, it is important to underline that the Lorentz-covariant kinematical space reduces to the usual $\SU(2)$ kinematical space of loop quantum gravity without making any gauge fixing. Besides, the $\SU(2)$ connection $\tilde{\mathcal{A}}$ does depend on the variable $\chi$. This reduction results from the crucial property that our unique Lorentz connection $\mathcal{A}$ lies in the conjugacy class of an $\SU(2)$ connection.

To finish this section, we are going to compute the action of the Lorentz-invariant area operator on the Lorentz-invariant kinematical states. We start by recalling the construction of the area operator. The variable $E^a_i$ can be smeared along a two-dimensional surface $S$ to give the electric flux
\be\nonumber
E_i(S)=\int_Sd^2x\,n_aE^a_i,
\ee
where $n_a$ is the normal to the surface.
At the classical level, the spatial metric is related to $E^a_i$ through the relation \ref{areaclassical}.
Notice that when the time-gauge is imposed, $\eta^{ij}$ becomes $\delta^{ij}$, and we recover the usual expression used in loop gravity. Therefore, if we divide the surface $S$ into $N$ cells, we can express its area as the regularized limit
\be\nonumber
\text{Area}(S)=\lim_{N\rightarrow+\infty}\sum_{I=1}^N\sqrt{E_i(S_I)\eta^{ij}E_j(S_I)}.
\ee
Because of what we have shown in the previous section, the surface operator is constructed in terms of the Casimirs of the Lorentz group, and is therefore Lorentz-invariant. 

Now we can act with this operator on the Lorentz kinematical states. More precisely, we act on an $\SU(2)$ spin-network state viewed as a function of $\tilde{\mathcal{A}}$. For simplicity, we will consider a simple graph consisting of a single link $\ell$ intersecting the elementary surface $S$ at a point $x$, in order to get only the contribution of the link $\ell$ to the area. Because of the property (\ref{holonomy}), the action of the surface operator on the holonomy of the connection $\mathcal{A}$ reduces to the action on $H_\ell(\tilde{\mathcal{A}})$. Moreover, since $E$ commutes with $\chi$, the operator $\hat{E}$ will only act on the first term in (\ref{su2 part}). It is then straightforward to show that we have
\be\nonumber
\left(\hat{E}_i(S)\eta^{ij}\hat{E}_j(S)\right)\triangleright H_\ell(\tilde{\mathcal{A}})=-\gamma^2\ell_\text{Pl}^2H_{\ell<x}(\tilde{\mathcal{A}})J_i(\mu\eta\mu)^{ij}J_jH_{\ell>x}(\tilde{\mathcal{A}}).
\ee
Then one can show that the matrix $(\mu\eta\mu)^{ij}$ is in fact the identity matrix! The expression above therefore reduces to
\be\nonumber
\left(\hat{E}_i(S)\eta^{ij}\hat{E}_j(S)\right)\triangleright H_\ell(\tilde{\mathcal{A}})=-\gamma^2\ell_\text{Pl}^2J^2H_\ell(\tilde{\mathcal{A}}),
\ee
where $J^2=J_iJ^i$ is the Casimir operator of $\SU(2)$. As a consequence, the contribution of the link $\ell$ colored with the representation $j$ on the Lorentz-covariant area, gives exactly the same expression as the one obtained in the framework of the time-gauge (\ref{area contribution}). In other words, the action of the Lorentz-invariant area operator on a Lorentz-invariant spin-network state is diagonal, and the spectrum is discrete and matches exactly that of standard loop quantum gravity. This result is highly nontrivial because one could have obtained a diagonal action of the area operator with a $\chi$-dependent spectrum. However, surprisingly, the dependence in $\chi$ disappears from the area spectrum.

\section*{Discussion and outlook}

\noindent In this paper, we have given the explicit construction and the proof of uniqueness of the unique commutative Lorentz-covariant connection for canonical gravity with nonvanishing Barbero-Immirzi parameter. When the time-gauge is not used, the canonical analysis of the Holst action features second-class constraints, which have to be handled by using either the Dirac bracket or an explicit solution. Following the work of Barros e S\'a \cite{barros}, we have used the parametrization of the phase space once the second-class constraints have been solved, to find suitable variables allowing to construct a commutative (in the sense of the Poisson bracket) one-form valued in $\sll(2,\mathbb{C})$, which transforms in a covariant way under the action of the Lorentz algebra. We have shown that it is possible to find a boost which sends this connection to a new connection with vanishing boost component. In other words, this proves that the Lorentz-covariant connection is gauge-equivalent to a pure $\su(2)$ connection. This property is crucial for the construction of the quantum theory in terms of usual $\SU(2)$ spin-network states. By applying the same construction to the electric field variable, we have been able to find a unique Lorentz-covariant electric field, up to trivial equivalence relations which are related to the symmetries of the equations of Lorentz covariance. Natural kinematical Lorentz invariant observables are therefore obtained computing the two Casimir functions associated to the electric field. It turns out that these two Casimir functions are in fact proportional to the area density, which gives a very interesting and very nice algebraic interpretation of the surface area. At the quantum level, we have finally shown that, surprisingly, the action of the Lorentz-invariant area operator on the holonomies of the Lorentz-covariant connection leads to the usual discrete $\SU(2)$ spectrum of time-gauge loop quantum gravity.

It is important to underline the fact that the Lorentz connection used in the present work is equivalent to the one that was found by Alexandrov earlier. However, because it makes use of the Dirac bracket, the construction of Alexandrov is technically more involved, less transparent, and, most importantly, the quantization cannot be performed rigorously. Using a solution to the second-class constraints enables one to work with a simple symplectic structure, and to derive the properties of the connection, the Lorentz-covariant electric field, and the surface operator. The claim that the time-gauge is not responsible for the discreteness of the geometric operators in loop quantum gravity is therefore clarified and understood with more details.

Many points are still to be understood. As a further development, it would be interesting to study for instance the volume operator. Indeed, there are several proposals for a regularization of the volume operator in the literature, and it might well be that the study of the Lorentz-covariant theory will discriminate between the different alternatives based on the nature of the spectrum. Even if the study of the area operator was rather simple, the volume operator being cubic in the electric field, it is \textit{a priori} nontrivial to see what will be its action on the Lorentz-covariant kinematical states. A more important point to study is the action of the covariant Hamiltonian constraint on the kinematical states. Indeed, the expression of this constraint is more complicated than it is in the time-gauge, but it might well be that its action on the covariant Lorentz connection reduces to a simple expression, which would simplify its analysis. Finally, this new look at Lorentz-covariant loop quantum gravity could provide a new way to understand the link between the canonical and spin-foam quantizations of gravity. It would be very nice to make contact with the work of Rovelli and Speziale \cite{rovelli-speziale}, who proposed a link between the kinematical boundary $\SU(2)$ theory and the Lorentz-invariant theory.

\subsection*{Aknowledgments}
\noindent We would like to thank Sergei Alexandrov for his comments on this work. It is also a pleasure to thank Etera Livine, Alejandro Perez, Carlo Rovelli, and Simone Speziale for their enthusiasm concerning this work. K. N. is partially supported by the ANR.

\appendix

\section{The Lorentz group and its algebra}

\noindent We recall in this appendix some basic definitions and notations relative to the Lorentz group $\SO(3,1)$ and its algebra $\so(3,1)$.

\subsection{The Lorentz algebra}
\label{lorentz algebra}

\noindent The Lorentz algebra $\so(3,1)$ is the Lie algebra of the isometry group $\SO(3,1)$ of the quadratic form $\eta=\text{diag}(-1,1,1,1)$. One of the basis of $\so(3,1)$ is defined by the three rotation generators $J_i$ and the three boost generators $P_i$ (with $i\in\{1,2,3\}$) satisfying the commutation relations
\be\nonumber
 [J_i,J_j]=\epsilon_{ij}^{~~k}J_k,\qquad\qquad
 [P_i,P_j]=-\epsilon_{ij}^{~~k}J_k,\qquad\qquad
 [P_i,J_j]=\epsilon_{ij}^{~~k}P_k,
\ee
where $\epsilon_{ijk}$ is the antisymmetric tensor defined by $\epsilon_{123}=1$. Here, indices are raised and lowered by the flat metric $\delta_{ij}$. The rotational algebra $\so(3)$ generated by $J_i$ ($i\in\{1,2,3\}$) forms a subalgebra of the Lorentz algebra.

The Lorentz algebra is of rank 2, and, therefore, its space of real symmetric invariant nondegenerate bilinear forms is of dimension 2. A basis of the space of nondegenerate bilinear forms is given by the two bilinear forms $\text{tr}_1(\cdot,\cdot)$ and
$\text{tr}_2(\cdot,\cdot)$ defined by 
\ba
&&\text{tr}_1(J_i,J_j)=\text{tr}_1(P_i,P_j)=0,\nonumber\\
&&\text{tr}_1(J_i,P_j)=\delta_{ij},\nonumber\\
&&\text{tr}_2(J_i,J_j)=-\text{tr}_2(P_i,P_j)=\delta_{ij},\nonumber\\
&&\text{tr}_2(J_i,P_j)=0.\nonumber
\ea
These two bilinear forms are canonically associated with two quadratic Casimir tensors $(C_1,C_2)\in\so(3,1)^{\otimes 2}$ defined by
\ba
 C_1&=&J_i\otimes P_i,\nonumber\\
 C_2&=&J_i\otimes J_i-P_i\otimes P_i.\nonumber
\ea
Here we have implicitly identified the Lie algebra $\so(3,1)$ with its dual. As usual, these two Casimir tensors are necessary to classify the irreducible representations of the Lie algebra.

Among the representations of $\so(3,1)$, two of them are useful to have a concrete intuition of what the Lorentz algebra is. These are the ``vectorial'' and the ``spinorial'' representations. None of them are unitary, because they are finite dimensional, but they are faithful, and therefore reproduce ``correctly'' the Lie algebra in terms of matrices.

The vectorial representation is four-dimensional. The element $\xi=u^iP_i+v^iJ_i\in\so(3,1)$, where $u$ and $v$ are vectors in $\mathbb R^3$, is represented by the matrix $\pi_v(\xi)$ defined by
\be\nonumber
\pi_v(\xi)=
\begin{pmatrix}
0 & u^1 & u^2 & u^3 \\
u^1 & 0 & -v^3 & v^2 \\
u^2 & v^3 & 0 & -v^1 \\
u^3 & -v^2 & v^1 & 0 \\
\end{pmatrix}
=
\begin{pmatrix}
 0 & u\tra \\
 u & \vb
\end{pmatrix}
\ee
The matrix on the right-hand side is a block matrix where $u$ denotes the vector $(u^1,u^2,u^3)$, $u\tra$ its transpose, and $\vb$ is the antisymmetric three-dimensional matrix canonically associated to the vector $v$, and defined by $\vb x=v\wedge x$ for any vector $x\in\mathbb R^3$. Note that the vectorial representation for a pure rotational element is reducible, and fundamentally the vectorial representation of the rotational algebra is three-dimensional. To any element $v^iJ_i$, one associates the antisymmetric three-dimensional matrix $\underline{v}$.

In the spinorial representation, the element $\xi\in\so(3,1)$ is represented by a complex two-dimensional traceless matrix $\pi_s(\xi)$. To define the spinorial representation,  it is convenient to introduce the Pauli matrices $\sigma_i$ defined by
\be\nonumber
\sigma_1=
\begin{pmatrix}
 1 & 0 \\
 0 & -1
\end{pmatrix},\qquad\qquad
\sigma_2=
\begin{pmatrix}
 0 & i \\
 -i & 0
\end{pmatrix},\qquad\qquad
\sigma_3=
\begin{pmatrix}
 0 & 1 \\
 1 & 0
\end{pmatrix}.
\ee
These matrices satisfy the relation $\sigma_i\sigma_j=\delta_{ij}+i\epsilon_{ij}^{~~k}\sigma_k$. As a consequence, we have
\be\nonumber
 \pi_s(\xi)=-\f{1}{2}(u^i+iv^i)\sigma_i.
\ee

\subsection{The Lorentz group}
\label{lorentz group}

\noindent The elements of the Lorentz group are obtained by exponentiation of the Lie algebra elements $\xi\in\so(3,1)$. As a result, in the vectorial and spinorial representations, Lorentz group elements are respectively four-dimensional and two-dimensional matrices. To write the group elements explicitly, it is simpler to distinguish between rotations and boosts.

A boost $B(u)$ is characterized by a vector $u\in\mathbb R^3$, or equivalently its direction $n\in\mathbb S^3$ and its norm $\theta=\sqrt{u^2}$. Its vectorial and spinorial representations are easy to obtain. They have the form
\ba
\pi_v\big(B(u)\big)&=&
\begin{pmatrix}
\ch &~~& n\tra\,\sh \\
n\,\sh &~~& 1+2\,\text{sh}^2\displaystyle\left(\f{\theta}{2}\right)n\tra\cdot n
\end{pmatrix}
=
\begin{pmatrix}
\ch &~~&  n\tra\,\sh \\
n\,\sh &~~& 1+(\ch-1)n\tra\cdot n
\end{pmatrix},\label{vectorial representation}\\
\pi_s\big(B(u)\big)&=&\text{ch}\left(\f{\theta}{2}\right)-\text{sh}\left(\f{\theta}{2}\right)\sigma_in^i.\label{spinorial representation}
\ea
Here, we use the notation $x\tra\cdot y$ to denote the matrix whose elements are $(x\tra\cdot y)_{ij}=x_iy_j$ for any pair of vectors $(x,y)$. 

A rotation $R(v)$ is also characterized by a vector $v\in\mathbb R^3$, or equivalently its direction $n\in\mathbb S^3$ and its norm $\theta=\sqrt{v^2}$. Its vectorial representation is three-dimensional and has the form
\be\nonumber
\pi_v\big(R(v)\big)=\cos\theta+(\sin\theta)\underline{n}+(1-\cos\theta)n\tra\cdot n.
\ee
The spinorial representation $\pi_s\big(R(v)\big)$ is easy to obtain from the spinorial representation $\pi_s\big(B(u)\big)$ of the boost, by simply replacing $\theta$ by $i\theta$ in the expression (\ref{spinorial representation}):
\be\nonumber
\pi_s\big(R(v)\big)=\cos\left(\f{\theta}{2}\right)-i\sin\left(\f{\theta}{2}\right)\sigma_in^i.
\ee

To finish, let us recall that any element of the Lorentz group can be written as the product $B(u)R(v)$ of a boost and a rotation. 

\subsection{Action of a boost}
\label{boost action}

\noindent The action of a boost $B(n,\theta)$, on a general connection $\mathcal{A}$ with boost and rotational components $\bst\mathcal{A}$ and $\rot\mathcal{A}$, gives a connection $\tilde{\mathcal{A}}$ with boost and rotational components $\bst\tilde{\mathcal{A}}$ and $\rot\tilde{\mathcal{A}}$. The transformation law is
\be\label{boost action}
\tilde{\mathcal{A}}\equiv B\triangleright\mathcal{A}=B^{-1}\mathcal{A}B+B^{-1}dB.
\ee 
Using for the boost $B$ the vectorial representation (\ref{vectorial representation}), we can calculate
\be\nonumber
B^{-1}=
\begin{pmatrix}
\ch &~~&  -n\tra\,\sh \\
-n\,\sh &~~& 1+(\ch-1)n\tra\cdot n
\end{pmatrix}
\ee
and
\be\nonumber
dB=
\begin{pmatrix}
\sh\,d\theta &~~&  n\tra\,\ch\,d\theta+dn\tra\,\sh \\
n\,\ch\,d\theta+dn\,\sh &~~& \sh\,d\theta(n\tra\cdot n)+(\ch-1)(dn\tra\cdot n+n\tra\cdot dn)
\end{pmatrix},
\ee
which gives
\be\nonumber
B^{-1}dB=
\begin{pmatrix}
0 &~~&  n\tra\,d\theta+dn\tra\,\sh \\
n\,d\theta+dn\,\sh &~~& (\ch-1)(dn\tra\cdot n-n\tra\cdot dn)
\end{pmatrix}.
\ee
One can then show that the action (\ref{boost action}) of the boost $B$ on the boost and rotation components $\brA$ of the connection is
\be\nonumber
B(n,\theta)\triangleright
\begin{pmatrix}
\bst\mathcal{A} \\
\rot\mathcal{A}
\end{pmatrix}
=
\begin{pmatrix}
\bst\tilde{\mathcal{A}} \\
\rot\tilde{\mathcal{A}}
\end{pmatrix}
=
\begin{pmatrix}
\bst\mathcal{A}\,\ch+(1-\ch)(n\cdot\bst\mathcal{A})n+(\rot\mathcal{A}\wedge n)\,\sh \\
(n\wedge\bst\mathcal{A})\,\sh+\rot\mathcal{A}+(\ch-1)n\wedge(\rot\mathcal{A}\wedge n)
\end{pmatrix}
+
\begin{pmatrix}
n\,d\theta+dn\,\sh \\
(\ch-1)n\wedge dn
\end{pmatrix}.
\ee

\section{The Lorentz algebra of constraints}
\label{appendixA}

\noindent In this appendix, we show that the boost and rotation generators satisfy the Lorentz algebra. To simplify the analysis, we introduce the decomposition
\ba
\mathcal{B}(u)&=&\mathcal{B}^{\text{\tiny{TG}}}(u)+\mathcal{B}^{\chi}(u)+\mathcal{B}^\text{\tiny{C}}(u)\nonumber\\
&=&-\int d^3x\:\partial_au\cdot E^a
+\int d^3x\:\Big[-u\cdot\zeta+(\zeta\cdot\chi)(\chi\cdot u)\Big]
+\int d^3x\left[\f{1}{\gamma}\partial_au\cdot(\chi\wedge E^a)-u\cdot(\chi\wedge E^a)\wedge A_a\right]\nonumber\\
&=&-\int d^3x\:\partial_au\cdot E^a
+\int d^3x\:\Big[-u\cdot\zeta+(\zeta\cdot\chi)(\chi\cdot u)\Big]
-\f{1}{\gamma}\int d^3x\:u\cdot\mathcal{D}_a^{(-\gamma A_a)}(\chi\wedge E^a),\nonumber\\
\mathcal{R}(v)&=&\mathcal{R}^{\text{\tiny{TG}}}(v)+\mathcal{R}^{\chi}(v)+\mathcal{R}^\text{\tiny{C}}(v)\nonumber\\
&=&\int d^3x\left[\f{1}{\gamma}\partial_av\cdot E^a+v\cdot(A_a\wedge E^a)\right]
+\int d^3x\:v\cdot(\chi\wedge\zeta)
+\int d^3x\:\partial_av\cdot(\chi\wedge E^a).\nonumber
\ea
The Poisson brackets between the components of the boost generator are
\ba
\big\{\mathcal{B}^{\text{\tiny{TG}}}(u_1),\mathcal{B}^{\text{\tiny{TG}}}(u_2)\big\}
&=&0,\nonumber\\
\big\{\mathcal{B}^{\chi}(u_1),\mathcal{B}^{\chi}(u_2)\big\}
&=&-\mathcal{R}^{\chi}(u_1\wedge u_2),\nonumber\\
\big\{\mathcal{B}^{\text{\tiny{C}}}(u_1),\mathcal{B}^{\text{\tiny{C}}}(u_2)\big\}
&=&\f{1}{\gamma}\int d^3x\:\Big[u_1\cdot(\partial_au_2\wedge\chi)\wedge(\chi\wedge E^a)-u_2\cdot(\partial_au_1\wedge\chi)\wedge(\chi\wedge E^a)\Big],\nonumber\\
%
\big\{\mathcal{B}^{\text{\tiny{TG}}}(u_1),\mathcal{B}^{\text{\tiny{C}}}(u_2)\big\}
&=&-\int d^3x\:u_2\cdot(\chi\wedge E^a)\wedge\partial_au_1,\nonumber\\
\big\{\mathcal{B}^{\chi}(u_1),\mathcal{B}^{\text{\tiny{C}}}(u_2)\big\}
&=&-\f{1}{\gamma}\int d^3x\:u_2\cdot\mathcal{D}_a^{(-\gamma A_a)}\Big[u_1\wedge E^a-(\chi\cdot u_1)(\chi\wedge E^a)\Big],\nonumber\\
\big\{\mathcal{B}^{\text{\tiny{TG}}}(u_1),\mathcal{B}^{\chi}(u_2)\big\}
&=&0,\nonumber\\
\big\{\mathcal{B}(u_1),\mathcal{B}(u_2)\big\}
&=&-\mathcal{R}^{\chi}(u_1\wedge u_2)-\mathcal{R}^{\text{\tiny{C}}}(u_1\wedge u_2)\nonumber\\
&&+\big\{\mathcal{B}^{\chi}(u_1),\mathcal{B}^{\text{\tiny{C}}}(u_2)\big\}
-\big\{\mathcal{B}^{\chi}(u_2),\mathcal{B}^{\text{\tiny{C}}}(u_1)\big\}
+\big\{\mathcal{B}^{\text{\tiny{C}}}(u_1),\mathcal{B}^{\text{\tiny{C}}}(u_2)\big\}\nonumber\\
&=&-\mathcal{R}(u_1\wedge u_2).\nonumber
\ea
For the rotation generator, we have
\ba
\big\{\mathcal{R}^{\text{\tiny{TG}}}(v_1),\mathcal{R}^{\text{\tiny{TG}}}(v_2)\big\}
&=&\mathcal{R}^{\text{\tiny{TG}}}(v_1\wedge v_2),\nonumber\\
\big\{\mathcal{R}^{\chi}(v_1),\mathcal{R}^{\chi}(v_2)\big\}
&=&\mathcal{R}^{\chi}(v_1\wedge v_2),\nonumber\\
\big\{\mathcal{R}^{\text{\tiny{C}}}(v_1),\mathcal{R}^{\text{\tiny{C}}}(v_2)\big\}
&=&0,\nonumber\\
\big\{\mathcal{R}^{\text{\tiny{TG}}}(v_1),\mathcal{R}^{\text{\tiny{C}}}(v_2)\big\}
&=&\int d^3x\:(E^a\wedge v_1)\cdot(\partial_av_2\wedge\chi),\nonumber\\
\big\{\mathcal{R}^{\chi}(v_1),\mathcal{R}^{\text{\tiny{C}}}(v_2)\big\}
&=&-\int d^3x\:(\chi\wedge v_1)\cdot(\partial_av_2\wedge E^a),\nonumber\\
\big\{\mathcal{R}^{\text{\tiny{TG}}}(v_1),\mathcal{R}^{\chi}(v_2)\big\}
&=&0,\nonumber\\
\big\{\mathcal{R}(v_1),\mathcal{R}(v_2)\big\}
&=&\mathcal{R}^{\text{\tiny{TG}}}(v_1\wedge v_2)
+\mathcal{R}^{\chi}(v_1\wedge v_2)
+\big\{\mathcal{R}^{\text{\tiny{TG}}}(v_1),\mathcal{R}^{\text{\tiny{C}}}(v_2)\big\}
+\big\{\mathcal{R}^{\chi}(v_1),\mathcal{R}^{\text{\tiny{C}}}(v_2)\big\}\cr\cr
&&-\big\{\mathcal{R}^{\text{\tiny{TG}}}(v_2),\mathcal{R}^{\text{\tiny{C}}}(v_1)\big\}
-\big\{\mathcal{R}^{\chi}(v_2),\mathcal{R}^{\text{\tiny{C}}}(v_1)\big\}\nonumber\\
&=&\mathcal{R}^{\text{\tiny{TG}}}(v_1\wedge v_2)
+\mathcal{R}^{\chi}(v_1\wedge v_2)\nonumber\\
&&+\int d^3x\:\Big[(E^a\wedge v_1)\cdot(\partial_av_2\wedge\chi)-(\chi\wedge v_1)\cdot(\partial_av_2\wedge E^a)\Big]\nonumber\\
&&-\int d^3x\:\Big[(E^a\wedge v_2)\cdot(\partial_av_1\wedge\chi)-(\chi\wedge v_2)\cdot(\partial_av_1\wedge E^a)\Big]\nonumber\\
&=&\mathcal{R}^{\text{\tiny{TG}}}(v_1\wedge v_2)
+\mathcal{R}^{\chi}(v_1\wedge v_2)\nonumber\\
&&+\int d^3x\:\Big[(v_1\wedge\partial_av_2)\cdot(\chi\wedge E^a)-(v_2\wedge\partial_av_1)\cdot(\chi\wedge E^a)\Big]\nonumber\\
&=&\mathcal{R}(v_1\wedge v_2).\nonumber
\ea
Finally, the Poisson brackets between the boosts and rotations are given by
\ba
\big\{\mathcal{B}^{\text{\tiny{TG}}}(u_1),\mathcal{R}^{\text{\tiny{TG}}}(v_2)\big\}
&=&\int d^3x\:\partial_au_1\cdot(E^a\wedge v_2),\nonumber\\
\big\{\mathcal{B}^{\text{\tiny{TG}}}(u_1),\mathcal{R}^{\chi}(v_2)\big\}
&=&0,\nonumber\\
\big\{\mathcal{B}^{\text{\tiny{TG}}}(u_1),\mathcal{R}^{\text{\tiny{C}}}(v_2)\big\}
&=&0,\nonumber\\
\big\{\mathcal{B}^{\chi}(u_1),\mathcal{R}^{\text{\tiny{TG}}}(v_2)\big\}
&=&0,\nonumber\\
\big\{\mathcal{B}^{\chi}(u_1),\mathcal{R}^{\chi}(v_2)\big\}
&=&\mathcal{B}^{\chi}(u_1\wedge v_2),\nonumber\\
\big\{\mathcal{B}^{\chi}(u_1),\mathcal{R}^{\text{\tiny{C}}}(v_2)\big\}
&=&\int d^3x\:(E^a\wedge\partial_av_2)\cdot(u_1-(\chi\cdot u_1)\chi),\nonumber\\
\big\{\mathcal{B}^{\text{\tiny{C}}}(u_1),\mathcal{R}^{\text{\tiny{TG}}}(v_2)\big\}
&=&\int d^3x\:\f{1}{\gamma}\Big[-(v_2\wedge E^a)\cdot(\chi\wedge\partial_au_1)+u_1\cdot\partial_av_2\wedge(\chi\wedge E^a)\Big]\cr\cr
&&+\int d^3x\:u_1\cdot\Big[(v_2\wedge A_a)\wedge(\chi\wedge E^a)+\Big((v_2\wedge E^a)\wedge\chi\Big)\wedge A_a\Big],\nonumber\\
\big\{\mathcal{B}^{\text{\tiny{C}}}(u_1),\mathcal{R}^{\chi}(v_2)\big\}
&=&\int d^3x\left[\f{1}{\gamma}(v_2\wedge\chi)\cdot(E^a\wedge\partial_au_1)-u_1\cdot\Big((v_2\wedge\chi)\wedge E^a\Big)\wedge A^a\right],\nonumber\\
\big\{\mathcal{B}^{\text{\tiny{C}}}(u_1),\mathcal{R}^{\text{\tiny{C}}}(v_2)\big\}
&=&\int d^3x\:(\chi\cdot u_1)\chi\cdot(E^a\wedge\partial_av_2),\nonumber\\
\big\{\mathcal{B}(u_1),\mathcal{R}(v_2)\big\}
&=&\mathcal{B}^{\text{\tiny{TG}}}(u_1\wedge v_2)+\mathcal{B}^{\chi}(u_1\wedge v_2)
+\big\{\mathcal{B}^{\text{\tiny{C}}}(u_1),\mathcal{R}^{\text{\tiny{TG}}}(v_2)\big\}
+\big\{\mathcal{B}^{\text{\tiny{C}}}(u_1),\mathcal{R}^{\chi}(v_2)\big\}\nonumber\\
&=&\mathcal{B}(u_1\wedge v_2).\nonumber
\ea

\section{Some useful formulas}
\label{formulas}

\noindent We give here some formulas that we have used in the core of the paper to compute the action of the boost and rotation generators on the general forms (\ref{generalform}) and (\ref{generalE}) of the connection and the momentum.

\subsection{Action of the boosts on the phase space variables}

\ba
\big\{\mathcal{B}(u),A\big\}&=&
du-\f{1}{\gamma}du\wedge\chi-(u\wedge A)\wedge\chi=du+\f{1}{\gamma T}Y\wedge du+\f{1}{T}Y\wedge(u\wedge A),\nonumber\\
\big\{\mathcal{B}(u),E\big\}&=&(\chi\wedge E)\wedge u=(\chi\cdot u)E-(E\cdot u)\chi,\nonumber\\
\big\{\mathcal{B}(u),\chi\big\}&=&u-(u\cdot\chi)\chi,\nonumber\\
\big\{\mathcal{B}(u),\zeta\big\}&=&-\f{1}{\gamma}du\wedge E+(A\wedge u)\wedge E+(\zeta\cdot\chi)u+\zeta(\chi\cdot u),\nonumber\\
\big\{\mathcal{B}(u),Y\big\}&=&Tu,\nonumber\\
\big\{\mathcal{B}(u),dY\big\}&=&Tdu+udT,\nonumber\\
\big\{\mathcal{B}(u),T\big\}&=&Y\cdot u,\nonumber\\
\big\{\mathcal{B}(u),f(T)\big\}&=&(Y\cdot u)f'(T),\nonumber\\
\big\{\mathcal{B}(u),Y\wedge A\big\}&=&\f{1}{\gamma T}\Big[(du\cdot Y)Y-Y^2du\Big]+Tu\wedge A+Y\wedge du+\f{1}{T}\Big[(A\cdot Y)(Y\wedge u)-(u\cdot Y)(Y\wedge A)\Big],\nonumber\\
\big\{\mathcal{B}(u),Y\wedge dY\big\}&=&Tu\wedge dY+dTY\wedge u+TY\wedge du,\nonumber\\
\big\{\mathcal{B}(u),(Y\cdot dY)Y\big\}&=&T(u\cdot dY)Y+dT(Y\cdot u)Y+T(Y\cdot du)Y+T(Y\cdot dY)u,\nonumber\\
\big\{\mathcal{B}(u),Y\wedge E\big\}&=&Tu\wedge E+Y\wedge(\chi\cdot u)E=Tu\wedge E+\f{1}{T}(Y\cdot u)Y\wedge E,\nonumber\\
\big\{\mathcal{B}(u),E\cdot Y\big\}&=&\f{1}{T}(Y\cdot u)(E\cdot Y)+\f{1}{T}E\cdot u,\nonumber\\
\big\{\mathcal{B}(u),(E\cdot Y)Y\big\}&=&\f{1}{T}(Y\cdot u)(E\cdot Y)Y+\f{1}{T}(E\cdot u)Y+(E\cdot Y)Tu,\nonumber
\ea

\subsection{Action of the rotations on the phase space variables}

\ba
\big\{\mathcal{R}(v),A\big\}&=&\chi\wedge dv-\f{1}{\gamma}dv+A\wedge v=\f{1}{T}\Yb dv-\f{1}{\gamma}dv-\vb A,\nonumber\\
\big\{\mathcal{R}(v),E\big\}&=&E\wedge v,\nonumber\\
\big\{\mathcal{R}(v),\chi\big\}&=&\chi\wedge v,\nonumber\\
\big\{\mathcal{R}(v),\zeta\big\}&=&E\wedge dv+\zeta\wedge v.\nonumber\\
\big\{\mathcal{R}(u),Y\big\}&=&Y\wedge v\nonumber\\
\big\{\mathcal{R}(u),dY\big\}&=&dY\wedge v+Y\wedge dv=-\vb dY+\Yb dv\nonumber\\
\big\{\mathcal{R}(u),T\big\}&=&0,\nonumber\\
\big\{\mathcal{R}(v),\Yb\big\}&=&[\Yb,\vb],\nonumber\\
\big\{\mathcal{R}(v),\Yb^2\big\}&=&[\Yb^2,\vb],\nonumber\\
\big\{\mathcal{R}(v),Y\wedge E\big\}&=&(Y\cdot v)E-(E\cdot v)Y,\nonumber\\
\big\{\mathcal{R}(v),E\cdot Y\big\}&=&0.\nonumber
\ea


\begin{thebibliography}{99}

\bibitem{al}
A. Ashtekar, J. Lewandowski,
Background Independent Quantum Gravity: A Status Report,
Class. Quant. Grav. \textbf{21}, R53 (2004).

\bibitem{rovelli}
C. Rovelli,
Quantum Gravity,
(Cambridge University Press, Cambridge, England, 2004).

\bibitem{thiemann} T. Thiemann,
Introduction to Modern Canonical Quantum General Relativity,
(Cambridge University Press, Cambridge, England, 2007).

\bibitem{sen} A. Sen,
Gravity as a spin system,
Phys. Lett. \textbf{B 119}, 89 (1982).

\bibitem{ashtekar-variables} A. Ashtekar,
New variables for classical and quantum gravity,
Phys. Rev. Lett. \textbf{57}, 2244 (1986).

\bibitem{barbero} J. F. Barbero,
Real Ashtekar variables for Lorentzian signature spacetimes,
Phys. Rev. \textbf{D 51} 5507 (1995).

\bibitem{immirzi} G. Immirzi,
Real and complex connections for canonical gravity,
Class. Quant. Grav. \textbf{14} L177 (1997), \texttt{arXiv:gr-qc/9612030}.

\bibitem{ashtekar-lewandowski} A. Ashtekar and J. Lewandowski,
Quantum theory of geometry I: Area operators,
Class. Quant. Grav. \textbf{14} A55 (1997), \texttt{arXiv:gr-qc/9602046}.

\bibitem{rovelli-smolin} C. Rovelli and L. Smolin,
Discreteness of area and volume in quantum gravity,
Nucl. Phys. \textbf{B 442} 593 (1995), \texttt{arXiv:gr-qc/9411005}.

\bibitem{rovelli-black hole} C. Rovelli,
Black hole entropy from loop quantum gravity,
Phys. Rev. Lett. \textbf{77} 3288 (1996), \texttt{arXiv:gr-qc/9603063}.

\bibitem{ABK} A. Ashtekar, J. Baez and K. Krasnov,
Quantum geometry of isolated horizons and black hole entropy,
Adv. Theor. Math. Phys. \textbf{4} 1 (2000), \texttt{arXiv:gr-qc/0005126}.

\bibitem{meissner} K. A. Meissner,
Black hole entropy in loop quantum gravity,
Class. Quant. Grav. \textbf{21} 5245 (2004), \texttt{arXiv:gr-qc/0407052}.

\bibitem{agullo} I. Agullo, J. F. Barbero, J. Diaz-Polo, E. Fernandez-Borja and E. J. S. Villase\~nor,
Black hole state counting in loop quantum gravity: A number theoretical approach,
Phys. Rev. Lett. \textbf{100} 211301 (2008), \texttt{arXiv:gr-qc/0005126}.

\bibitem{ENP} J. Engle, K. Noui and A. Perez,
Black hole entropy and SU(2) Chern-Simons theory,
Phys. Rev. Lett. \textbf{105} 031302 (2010), \texttt{arXiv:0905.3168 [gr-qc]}.

\bibitem{rovelli-thiemann} C. Rovelli and T. Thiemann,
The Immirzi parameter in quantum general relativity,
Phys. Rev. \textbf{D 57} 1009 (1998), \texttt{arXiv:gr-qc/9705059}.

\bibitem{samuel} J. Samuel,
Is Barbero's Hamiltonian formulation a gauge theory of Lorentzian gravity?,
Class. Quant. Grav. \textbf{17} L141 (2000), \texttt{arXiv:gr-qc/0005095}.

\bibitem{alexandrov1} S. Alexandrov and D. V. Vassilevich,
Path integral for the Hilbert-Palatini and Ashtekar gravity,
Phys. Rev. \textbf{D 58} 124029 (1998), \texttt{arXiv:gr-qc/9806001}.

\bibitem{alexandrov2} S. Alexandrov,
SO(4,C)-covariant Ashtekar-Barbero gravity and the Immirzi parameter,
Class. Quant. Grav. \textbf{17} 4255 (2000), \texttt{arXiv:gr-qc/0005085}.

\bibitem{alexandrov3} S. Alexandrov and D. V. Vassilevich,
Area spectrum in Lorentz-covariant loop gravity,
Phys. Rev. \textbf{D 64} 044023 (2001), \texttt{arXiv:gr-qc/0103105}.

\bibitem{alexandrov4} S. Alexandrov and E. R. Livine,
SU(2) loop quantum gravity seen from covariant theory,
Phys. Rev. \textbf{D 67} 044009 (2003), \texttt{arXiv:gr-qc/0209105}.

\bibitem{holst} S. Holst,
Barbero's Hamiltonian derived from a generalized Hilbert-Palatini action,
Phys. Rev. \textbf{D 53} 5966 (1996), \texttt{arXiv:gr-qc/9511026}.

\bibitem{henneaux-teitelboim} M. Henneaux and C. Teitelboim,
Quantization of gauge systems,
(Princeton University Press, Princeton, 1994).

\bibitem{cianfrani-montani} F. Cianfrani and G. Montani,
Towards loop quantum gravity without the time-gauge,
Phys. Rev. Lett. \textbf{102}, 091301 (2009), \texttt{arXiv:0811.1916 [gr-qc]}.

\bibitem{barros} N. Barros e S\'a,
Hamiltonian analysis of general relativity with the Immirzi parameter,
Int. J. Mod. Phys. \textbf{D 10} 261 (2001), \texttt{arXiv:gr-qc/0006013}.

\bibitem{alexandrov5} S. Alexandrov,
On choice of connection in loop quantum gravity,
Phys. Rev. \textbf{D 65} 024011 (2001), \texttt{arXiv:gr-qc/0107071}.

\bibitem{GLNS} M. Geiller, M. Lachi\`eze-Rey, K. Noui and F. Sardelli,
A Lorentz-covariant connection for canonical gravity,
Submitted to SIGMA Special Issue ``Loop Quantum Gravity and Cosmology'', \texttt{arXiv:1103.4057 [gr-qc]}.

\bibitem{peldan} P. Peldan,
Actions for gravity, with generalizations: A review,
Class. Quant. Grav. \textbf{11} 1087 (1994), \texttt{arXiv:gr-qc/9305011}.

\bibitem{lost} J. Lewandowski, A. Okolow, H. Sahlmann and T. Thiemann,
Uniqueness of diffeomorphism invariant states on holonomy-flux algebras,
Commun. Math. Phys. \textbf{267} 703 (2006), \texttt{arXiv:gr-qc/0504147}.

\bibitem{freidel-livine} L. Freidel and E. R. Livine,
Spin networks for non-compact groups,
J. Math. Phys. \textbf{44} 1322 (2003), \texttt{arXiv:hep-th/0205268}.

\bibitem{rovelli-speziale} C. Rovelli and S. Speziale,
Lorentz covariance of loop quantum gravity,
Phys. Rev. \textbf{D 83} 104029 (2011), \texttt{arXiv:1012.1739 [gr-qc]}.

\end{thebibliography}
\end{document}